\documentclass[twocolumn,prb,superscriptaddress,showpacs,preprintnumbers,amsmath,amssymb]{revtex4}

\usepackage{graphicx}
\usepackage{dcolumn}
\usepackage{bm}

\renewcommand{\Re}{\text{Re}}
\renewcommand{\Im}{\text{Im}}

\begin{document}

\title{Quantum transport theory of anomalous electric, thermoelectric, and thermal Hall effects in ferromagnets}

\author{Shigeki Onoda}
 \email{s.onoda@riken.jp}
\affiliation{%
RIKEN (The Institute of Physical and Chemical Research), 2-1, Hirosawa, Wako 351-0198, Japan
}%
\author{Naoyuki Sugimoto}%
\affiliation{
  Department of Applied Physics, University of Tokyo, Tokyo 113-8656, Japan
}%
\author{Naoto Nagaosa}
\affiliation{
  CREST, Department of Applied Physics, University of Tokyo, 7-3-1, Hongo, Tokyo 113-8656, Japan}
\date{\today}

\begin{abstract}
  In ferromagnets, the spontaneous magnetization bears the Hall effect through the relativistic spin-orbit interaction. Similar effects also occur in thermoelectric and thermal transport phenomena. Their mechanism, if it is of the intrinsic or extrinsic origin, has been controversial for many decades. We present a unified theory of these Hall transport phenomena in ferromagnetic metals with dilute impurities at the zero temperature, in terms of a fully quantum-mechanical transport theory for multi-band systems with the self-consistent $T$-matrix approximation. This theory becomes exact with a single impurity and is appropriate for treating the dilute limit of the impurity concentration $n_{\text{imp}}$.
With the Fermi energy $E_F$ and the spin-orbit interaction energy $E_{\text{SO}}$ being fixed ($E_F>E_{\text{SO}}$), three regimes and the associated two crossovers are found in the anomalous Hall conductivity $\sigma_{xy}$ as a function of $n_{\text{imp}}$ that controls the longitudinal conductivity $\sigma_{xx}$. (i) In the superclean case with the relaxation rate $\hbar/\tau \lesssim u_{\text{imp}}E_{\text{SO}}D$, the skew scattering arising from the vertex correction yields a dominant contribution that is inversely proportional to $n_{\text{imp}}$, where $u_{\text{imp}}$ is the impurity potential strength and $D$ is the density of states. With increasing $\hbar/\tau$, this extrinsic skew-scattering contribution rapidly decays. (ii) In the moderately dirty regime $u_{\text{imp}}E_{\text{SO}}D \lesssim \hbar/\tau \lesssim E_F$, $\sigma_{xy}$ becomes insensitive to the scattering strength because of the intrinsic dissipationless topological Berry-phase contribution. It is resonantly enhanced to the order of the quantization unit of conductance, when an accidental degeneracy of band dispersions around the Fermi level is lifted by the spin-orbit interaction. Further increasing $\hbar/\tau$, another crossover occurs to (iii) the scaling regime of $\sigma_{xy}\propto\sigma_{xx}^\varphi$ with $\varphi\sim1.6$, which has recently been verified by experiments on a wide class of ferromagnets. Similar behaviors also appear in the temperature-linear coefficient of the themal Hall conductivity $\kappa_{xy}$. The thermoelectric Hall conductivity $\alpha_{xy}$ strongly diverges in the clean limit when the Fermi level crosses edges of the avoided-crossing, which may be observed by careful experiments. With increasing $\hbar/\tau$, there occurs an interference beteween positive and negative contributions to $\alpha_{xy}$, which often leads to a sign change and obscures similar crossovers in the anomalous Nernst effect.
\end{abstract}

\maketitle

\section{Introduction}\label{sec:intro}

The Hall effect is a fundamental transport phenomenon in solids that an applied electric current induces a transverse voltage drop, or an applied electric field produces a transverse current~\cite{Hall1879,Hall1880,Hurd}. In conventional semiconductors and metals, this Hall current linear in a weak magnetic field $H$ offers a mean to probe an effective carrier number through the normal Hall coefficient $R_H$. In contrast to this normal Hall effect driven by the Lorentz force, a spontaneous magnetization as well bears the Hall effect in ferromagnets~\cite{Hall1880}, where the relativistic spin-orbit interaction is indispensable for connecting the spin polarization with the orbital motion of electrons. This phenomenon, i.e., {\it spontaneous or anomalous Hall effect}~\cite{Hall1880,Hurd}, has been one of the most fundamental and intriguing issues in condensed-matter physics. 

Early experimental works led an empirical relation of the Hall resistivity $\rho_{xy}$ to the {\it weak} applied magnetic field $H^z$ and the spontaneous magnetization $M^z$ both along the $z$ direction; 
\begin{equation}
  \rho_{xy} \approx R_H H^z + 4 \pi R_s M^z
  \label{eq:rho_xy}
\end{equation}
with $R_s$ being called the anomalous Hall coefficient~\cite{Hurd}. Similar spontaneous or anomalous effects are also found in thermoelectric and thermal transport phenomena as the anomalous Nernst-Ettingshausen effect and the anomalous Luduc-righi effect, respectively. In spite of the intensive and extensive studies for many decades~\cite{KarplusLuttinger54,Smit55,Smit58,Luttinger58,Kondo62,Fukuyama,Berger70,Berger72,Nozieres73,Coleman85}, a long standing debate on the mechanism has not been resolved yet. The keen issues are roles of scattering and the associated relaxation and dissipation. In particular, the intrinsic vs. extrinsic mechanisms and the associated scaling behaviors~\cite{Onoda06_prl} of the anomalous Hall effect have attracted revived interest because of the fundamental importance of the dissipationless and topological nature of the intrinsic mechanism, which penetrates the whole debates on this issue~\cite{KarplusLuttinger54,Smit55,Smit58,Luttinger58,Kondo62,Fukuyama,Berger70,Berger72,Nozieres73,Coleman85,Onoda06_prl,Ohno,Taguchi01,Lee_science04,Manyala_nma04,Ong07,Miyasato07}.  

In a recent Letter~\cite{Onoda06_prl}, we have presented a unified theory of the anomalous Hall effect, fully taking account of both the intrinsic and the extrinsic contributions on an equal footing. Now, main aims of this paper are (i) to provide a comprehensive description of the unified theory of the anomalous electric, thermoelectric, and thermal Hall transport coefficients with some important details of the formalism and the calculation procedure, (ii) to explain key experimental observations in various ferromagnetic metals, including  the magnitudes of $\sigma_{xx}$ and $\sigma_{xy}$ and their scaling relations, by means of the classification into three regimes revealed by the theory, and accordingly (iii) to resolve the long standing controversy on the mechanism.

The dissipationless and topological nature involved in the Hall effect has been highlighted by the discovery of quantum Hall effect~\cite{Prange} in two-dimensional ($d=2$) disordered electron systems under a {\it strong} magnetic field. For the Bloch electrons in perfect crystal, the Hall conductivity is expressed by the Thouless-Kohmoto-Nightingale-Nijs (TKNN) formula~\cite{TKNN},
\begin{equation}
  \sigma_{ij}^{\text{TKNN}} = -\epsilon_{ij\ell}e^2\hbar
  \sum_n\int\!\frac{d^d\bm{p}}{(2\pi\hbar)^d}b_n^\ell(\bm{p}) f(\varepsilon_n(\bm{p}))
  \label{eq:TKNN}
\end{equation}
with the electronic charge $-e$ ($e>0$), the Planck constant $h=2\pi\hbar$, the Fermi distribution function $f(\varepsilon)$, and the anti-symmetric tensor $\epsilon_{ij\ell}$. We have introduced the eigenenergy $\varepsilon_n(\bm{p})$, the Berry-phase connection 
\begin{equation}
  \bm{a}_n(\bm{p}) = i \langle n,{\bm{p}}| \bm{\nabla}_p | n,{\bm{p}}\rangle,
  \label{eq:a}
\end{equation}
and the Berry-phase curvature 
\begin{equation}
  \bm{b}_n(\bm{p}) = \bm{\nabla}_p\times \bm{a}_n(\bm{p})
  \label{eq:b}
\end{equation}
of the generalized Bloch wave function $|n,\bm{p}\rangle$ with the band index $n$ and the Bloch momentum $\bm{p}$. Each band is characterized by a topological integer called the Chern number 
\begin{equation}
  C_n\equiv -\int\frac{dp_x p_y}{(2\pi)^2}b_n^z(\bm{p}). 
  \label{eq:C_n}
\end{equation}
The sum of $C_n$ over the occupied bands determines the integer $\nu$ (Chern number) for the quantization of the Hall conductivity $\sigma_{xy}=\nu e^2/h$. Then, in ideal cases when the Fermi level is located within an energy gap, the longitudinal conductivity $\sigma_{xx}$ vanishes and the Hall conductivity $\sigma_{xy}$ is quantized in a unit of $e^2/h=3.87\times10^{-5}\ \Omega^{-1}$. This Berry-phase effect has been incorporated into the adiabatic semi-classical wave-packet equations for the Boltzmann transport theory~\cite{SundaramNiu99}.

Historically, the dissipationless thermodynamic Hall current was first discussed by Karplus-Luttinger~\cite{KarplusLuttinger54}. They initiated an intrinsic mechanism of the anomalous Hall effect in a band model for ferromagnetic metals with the spin-orbit interaction. Recognizing that the inter-band matrix element of the current operator plays a key role, they derived a generic expression for the band-intrinsic contribution to the anomalous Hall conductivity, which is independent of the scattering rate. This accounts for the experimentally observed scaling relation for the resistivity tensor $\rho_{ij}$, i.e., $\rho_{xy}\propto\rho_{xx}^2$ or equivalently, $\sigma_{xy}$ being constant. They also performed a perturbation expansion in $M$ and the spin-orbit coupling $\xi$ to derive the empirical law given by Eq.~(\ref{eq:rho_xy}). However, there are two drawbacks in the theory.

\begin{figure}
  \begin{center}
    \includegraphics[width=8.0cm]{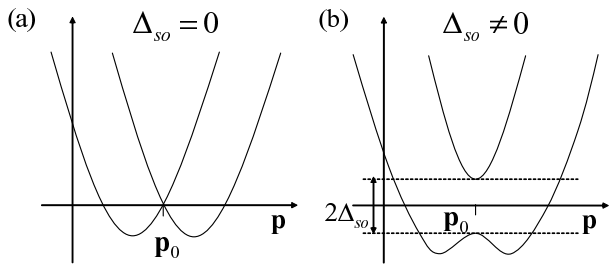}
  \end{center}
  \caption{ The simplest examples of (a) an accidental crossing of two band dispersions at a momentum $\textbf{p}_0$ and (b) an avoided-crossing with a splitting of the dispersions by $2E_{\text{SO}}$ at this momentum region.}
  \label{fig:crossing}
\end{figure}

Firstly, the perturbation theory for $\sigma_{xy}$ in terms of the spin-orbit interaction energy $E_{\text{SO}}\sim\xi M$, which is usually even small compared with the bandwidth or the Fermi energy $E_F$ except in the $f$ electrons, can not capture a topological nature involved in the intrinsic anomalous Hall effect. Recently, it has been recognized that the Karplus-Luttinger's general expression for the band-intrinsic contribution actually coincides with the TKNN formula given by Eq.~(\ref{eq:TKNN})~\cite{MOnodaNagaosa02,JungwirthNiuMacDonald02}, and that each band contains a finite Chern number~\cite{MOnodaNagaosa02}, as in the quantum Hall systems. Without the spin-orbit interaction, the Hamiltonians describing the majority and minority spin bands are decoupled in the band theory. Then, the fact that the Hamiltonians and the Bloch wavefunctions are real requires the accidental degeneracy of band dispersions in the three-dimensional Brillouin zone~\cite{Herring2}. Turning on the spin-orbit interaction, this condition no longer holds and then the accidental band crossings are avoided, leaving a small energy separation of the order of $E_{\text{SO}}$. Namely, the spin-orbit interaction plays a crucial role in avoiding a crossing of band dispersions at a certain momentum $\bm{p}_0$ (see Fig.~\ref{fig:crossing}). This avoided-crossing of band dispersions is accompanied by a transfer of Chern numbers among the two-dimensional bands. This phenomenon called ``parity anomaly'' in $(2+1)$ dimensions has a non-perturbative nature~\cite{parity}: $\sigma_{xy}$ exhibits a discontinuous jump by $e^2/h$ as $E_{SO}$ continuously changes its sign. This points to an importance of the avoided-crossing of band dispersions near the chemical potential. Therefore, the nontrivial topological structure in the Bloch wave functions of ferromagnets is not captured by the perturbative treatment~\cite{KarplusLuttinger54} of the spin-orbit coupling $\xi$ leading to the empirical law given by Eq.~(\ref{eq:rho_xy}) with $R_s\propto\xi$.

This picture based on the ``parity anomally'' has been supported by recent first-principles calculations. When the Fermi level is located around such an avoided-crossing of dispersions, as found in recent {\it ab inito} calculations for SrRuO$_3$~\cite{Fang03}, the bcc Fe~\cite{Yao04,Vanderbilt06}, CuCr$_2$Se$_{4-x}$Br$_x$~\cite{YaoFang07}, Co~\cite{Vanderbilt07}, and Ni~\cite{Vanderbilt07}, the magnitude of $\sigma_{xy}^{\text{TKNN}}$ is resonantly enhanced; $\sigma_{xy}\sim e^2/ha\sim 10^3 \ \Omega^{-1}\ {\rm cm}^{-1}$ with the lattice constant $a\approx4$~\AA~\cite{Fang03,Yao04}, which can be regarded as an nearly quantized $\sigma_{xy}$ in each two-dimensional momentum plane or $p_z$. This resonant enhancement of $\sigma_{xy}$ without any small factor of $E_{\text{SO}}\sim\xi M$ means that the perturbation expansion in $\xi M$ fails when the Fermi level is located within the energy range of the avoided-crossing of band dispersions. In the metallic system, there appear many avoided-crossings and/or more complex structures near the Fermi level, which may lead to a complex behavior of $\sigma_{xy}$ as a function of the chemical potential, the magnetization, and the crystal structure analogous to the quantum chaos, as actually found in first-principles calculations~\cite{Fang03,Yao04,YaoFang07}. Interference among the contributions from different bands and/or different momentum regions may often reduces the magnitude of $\sigma_{xy}$, but its variation is of the order of $e^2/ha$ in both calculations and experiments on SrRuO$_3$~\cite{Fang03}. 

Secondly, scattering events extrinsic to the band structure were completely ignored in the Karplus-Luttinger's theory~\cite{KarplusLuttinger54}. In fact, adiabatic semi-classical Boltzmann transport analyses~\cite{Smit55,Smit58,Kondo62,Berger70,Berger72,Nozieres73,Coleman85}, which have also been taken over to the extrinsic scenario for the spin Hall effect~\cite{EngelHalperinRashba05}, have revealed that the spin-orbit interaction in the impurity potential produces the anomalous Hall effect through the skew scattering or equivalently the Mott scattering~\cite{Smit55,Smit58,Luttinger58,Nozieres73} and the side jump~\cite{Berger70,Berger72,Nozieres73}. The skew-scattering contribution diverges in the clean limit ($\sigma_{xx}\sim (e^2/ha)(E_F\tau/\hbar)\to\infty$) as
\begin{equation}
  \sigma_{xy}^{\text{skew}}=S\sigma_{xx}.
\label{eq:skew}
\end{equation}
Here, $S\sim E_{\text{SO}}u_{\text{imp}}D/E_F$ ($|S|\ll1$) is the skewness factor with the density of states $D$, the Fermi energy $E_F$, and the impurity potential strength $u_{\text{imp}}$, respectively. Accordingly, Luttinger reconsidered the issue by means of the expansions of $\sigma_{xy}$ in $u_{\text{imp}}$~\cite{KohnLuttinger57,Luttinger58} and the impurity concentration $n_{\text{imp}}$~\cite{LuttingerKohn58}, or in $\hbar/(E_F\tau)$ with the relaxation time $\tau\sim\hbar/n_{\text{imp}}u_{\text{imp}}^2D$. Then, the leading-order term is proportional to $1/(n_{\text{imp}}u_{\text{imp}})$, which corresponds to the skew-scattering contribution~\cite{Luttinger58}. The subleading-order term, which is of the zeroth order in $u_{\text{imp}}$ and $n_{\text{imp}}$, includes the original Karplus-Luttinger's result~\cite{KarplusLuttinger54} as well as some other terms that partiallly cancel the intrinsic contribution. The side-jump contribution has the form~\cite{Nozieres73}
\begin{equation}
  \sigma_{xy}^{\text{sj}}=2n_{\text{el}}e^2\lambda M^z
  \label{eq:side-jump}
\end{equation}
in the clean limit, with the electron density $n_{\text{el}}$ and the relativistic Aharonov-Cacher coupling $\lambda$ leading to an energy shift by $\lambda\bm{p}\times\bm{M}\cdot\bm{E}$ due to the applied electric field $\bm{E}$. It is remarkable that this side-jump contribution is insensitive to the relaxation rate, leading to the scaling relation $\rho_{xy}\propto\rho_{xx}^2$. Therefore, it can be incorporated into the subleading term in the Luttinger's expansion of $\sigma_{xy}$.

In the conventional quantum transport theory given by Luttinger~\cite{Luttinger58}, the anomalous Hall conductivity is expanded in $E_{\text{SO}}$. Then, the ratio $\hbar/(E_F \tau)$ is the only key expansion parameter. The sum of the leading skew-scattering contribution and the sub-leading intirnsic and other impurity-independent contributions to $\sigma_{xy}$ is given by
\begin{equation}
  \sigma_{xy} \sim { {e^2} \over {h a}} \left[
  S {{  E_F \tau} \over {\hbar} }  + c E_{\text{SO}}D
  + \cdots \right]
  \label{eq:Luttinger1}
\end{equation}
with $c$ being a constant of the order of unity. Assuming $u_{\text{imp}}\cong E_F$, the ratio $E_{\text{SO}}D$ appears only as the overall factor;
\begin{equation}
  \sigma_{xy}\sim\frac{e^2}{ha}E_{\text{SO}}D
  \left[\frac{E_F\tau}{\hbar}+c+\cdots\right].
  \label{eq:Luttinger2}
\end{equation}
In this expression, $\hbar/(E_F \tau)$ is the only relevant parameter that controls an extrinsic-intrinsic crossover. Namely, as far as the expansion in $E_{\text{SO}}D$ is valid, the first term in Eq.~(\ref{eq:Luttinger1}) as a skew-scattering contribution is dominant over the other terms in the clean metal $\hbar/(E_F \tau)\ll 1$. Therefore, it has been believed that the extrinsic skew-scattering mechanism is dominant~\cite{Hurd}, as supported by some experiments around the ferromagnetic Curie temperature~\cite{Kondo62} and in heavy-fermion compounds showing a large susceptibility~\cite{Maranzana67,Coleman85}.

Nevertheless, many experimental and theoretical works support the Karplus-Luttinger's scenario. At a fixed impurity potential strength $u_{\text{imp}}$, experimental results on Fe- and Ni-based dilute alloys~\cite{Hurd}, CuCr$_2$Se$_{4-x}$Br$_x$~\cite{Lee_science04,Miyasato07}, and semiconducting helimagnets Fe$_{1-y}$Co$_y$Si and Fe$_{1-y}$Mn$_y$Si~\cite{Manyala_nma04} appeared to be consistent with the Karplus-Luttinger's prediction $\rho_{xy} \propto \rho_{xx}^2$. First-principles calculations of the anomalous Hall conductivity for SrRuO$_3$~\cite{Fang03} and Fe~\cite{Yao04} in terms of the Karplus-Luttinger's scenario or equivalently the TKNN formula~\cite{TKNN} given by Eq.~(\ref{eq:TKNN}) also show large values of $\sigma_{xy}\sim e^2/ha$ which agree with the experimentally observed values at low temperatures. The agreement is not consistent with a simple-minded perturbation expantion of the intrinsic contribution in $\xi M$, namely, the second term in the square bracket of Eq.~(\ref{eq:Luttinger1}) or (\ref{eq:Luttinger2}). This points to the importance of the resonat enhancement due to the topological nature and urges a reexamination of the intrinsic mechanism against the extrinsic one. Theoretically, the two characters involved in the anomalous Hall effect have not been seriously considered on an equal footing. A unified description of both intrinsic and extrinsic contributions is called for. It is also helpful to develop a theory of the anomalous thermoelectric Hall (Nernst-Eittingshausen) effect and the anomalous thermal Hall (Leduc-Righi) effect in the same theoretical framework. To provide a comprehensive description of the unified theory on these anomalous Hall transport phenomena is the main scope of the present study.

This paper is organized as follows: In Sec.~\ref{sec:model}, we introduce a model appropriate for studying the interplay between the topological dissipationless Hall current and the extrinsic scattering events. In Sec.~\ref{sec:theory}, the Keldysh Green's function formalism in the gauge-covariant Wigner space is briefly explained, together with the self-consistent $T$-matrix approximation, the numerical results for the equilibrium properties, the Mott rule, and the Wiedemann-Franz law. Then, In Sec.~\ref{sec:AHE}, numerical results are given for the anomalous electric  and thermal Hall conductivities, and the scaling relation between $\sigma_{xx}$ and $\sigma_{xy}$. In Sec.~\ref{sec:ANE}, using the Mott rule, we show numerical results of the thermoelecrtic Hall conductivity for the anomalous Nernst effect. In Sec.~\ref{sec:others}, we clarify relations of the present theory to other theories. In Sec.~\ref{sec:exp}, experimental results on $\sigma_{xy}$ and $\sigma_{xx}$ are summarized and compared with the present theory. Conclusions are drawn in Sec.~\ref{sec:conclusions}. Some necessary details of calculations of the Green's functions and the self-energy in our formalism are given in Appendices.

\section{Model}\label{sec:model}

\begin{figure}[htb]
  \begin{center}
    \includegraphics[width=7.2cm]{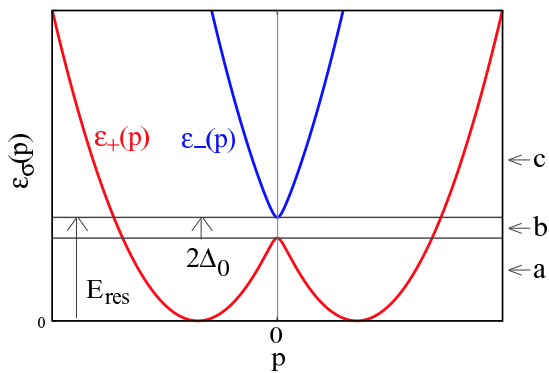}
  \end{center}
  \caption{(Color online) Electronic band dispersions of the present model given by $\hat{H}_0$. }
\label{fig:dispersion}
\end{figure}

A simple model that fully takes into account both the ``parity anomaly'' associated with the avoided-crossing of band dispersions and the impurity scattering can be obtained by expanding the Hamiltonian at a fixed $p_z$ with respect to the momentum $\bm{p}$ measured from the originally crossing point $\bm{p}_0$ of two dispersions;
\begin{subequations}
  \begin{eqnarray}
    \hat{H} &=& \hat{H}_0+\hat{H}_\text{imp},
    \label{eq:H}\\
    \hat{H}_0(\bm{p}) &=&-\Delta_0\hat{\sigma}^z+v\bm{p}\cdot\hat{\bm{\sigma}}\times\bm{e}^z+\frac{\bm{p}^2}{2m}\hat{\sigma}^0
    \label{eq:H_0}\\
    \hat{H}_\text{imp} &=& u_{\text{imp}}\hat{\sigma}^0\sum_{\bm{r}_{\rm imp}}\delta(\bm{r}-\bm{r}_{\rm imp})
    \label{eq:H_imp}
  \end{eqnarray}
  \label{eq:Hs}
\end{subequations}
with the position $\bm{r}$ of electron, the Pauli and identity matrices $\hat{\bm{\sigma}}=(\hat{\sigma}_x,\hat{\sigma}_y,\hat{\sigma}_z)$ and $\hat{\sigma}^0$, respectively, and the unit vector $\bm{e}^z$ in the $z$ direction. The first term corresponds to the level splitting $2\Delta_0=E_{\text{SO}}$ of two bands at the avoided-crossing momentum. The second term gives the linear dispersion with the velocity $v$. The third term represents the quadratic dispersion with an effective mass $m$, whose anisotropy has been neglected since it is unimportant. When $\Delta_0=0$, $\hat{H}_0(\bm{p})$ given by Eq.~(\ref{eq:H_0}) is reduced to the two-dimensional Rashba model~\cite{BychkovRashba84} for hetrostructures of $n$-type semiconductors, which possesses a magnetic monopole at $\bm{p}=\bm{0}$. However, note the significant difference in the physical interpretation of the model parameters. In the present model for ferromagnetic metals, $\Delta_0$ plays the role of the spin-orbit interaction that lifts the accidental degeneracy of two band dispersions with the velocity $v$, while for semiconductors, $v$ is the spin-orbit coupling and $\Delta$ is the Zeeman splitting.

For simplicity, we also assume that the impurity potential has the $\delta$-functional form of the strength $u_{\text{imp}}$ at random positions $\{\bm{r}_{\rm imp}\}$, since as we will show, this simple form bears the extrinsic skew-scattering and side-jump contributions. The generalization to long-range potential is straitforward but requires rather lengthy calculations. Here, note the sgnificant difference in the way of taking the impurity average. One might think that the conventional scheme of taking the average over the impurity positions, which we adopt, might be mimicked by taking the potential average in the white noise model or the random potential model. However, this is true only within the Born approximation. In fact, the higher-order scattering proccesses, which are crucial for the skew-scattering contribution, can not be correctly reproduced by the random potential model.

Frist let us consider the model in the absence of the impurities. This model has two band dispersions $\varepsilon_\pm(p)$ separated by $2\Delta_p$ with $\Delta_p=\sqrt{v^2 p^2+\Delta_0^2}$, as shown in Fig.~\ref{fig:dispersion}. Henceforth, the bottom of the lower band in the absence of impurities is chosen as the origin of the energy and the bottom of the upper band denoted as $E_{\rm res}=\varepsilon_-(p=0)$ is taken as an energy unit. 
The present model possesses the gauge flux
\begin{equation}
  b^z_{\sigma,\bm{p}} =-\sigma { {v^2\Delta_0} 
    \over { 2\Delta_p^3}}
  \label{eq:b_H_0}
\end{equation}
 for the band index $\sigma=\pm$, as discussed in the literature~\cite{Dugaev05}. Substituting Eq.~(\ref{eq:b_H_0}) into Eq.~(\ref{eq:TKNN})  and the integration over the momentum yields the intrinsic contribution to the anomalous Hall conductivity,
\begin{equation}
  \sigma_{xy}^{\text{TKNN}}=-\frac{e^2}{2h}\sum_\sigma\sigma\frac{\Delta_0}{\Delta_{p_\sigma}},
  \label{eq:sigma_xy_H_0}
\end{equation}
when the both bands are partially occupied. Here, $p_\sigma$ is the Fermi momentum for the band index $\sigma$.

When $E_F \in [E_{\rm res}-2\Delta_0,E_{\rm res}]$, $\sigma_{xy}^{\text{TKNN}}$ is resonantly enhanced and approaches the maximum value $e^2/2h$. Away from this resonance, dominant contributions from the momentum region around $\bm{p}=0$ cancel out each other or do not appear, leading to a suppression of $\sigma_{xy}^{\text{TKNN}}(\approx(e^2/h)(E_{\text{SO}}/E_F))$, and then the perturbation expansion in $E_{\text{SO}}$ is justified. 
Therefore, the present model, Eq.~(\ref{eq:Hs}), can be regarded as a minimal continuum model for a momentum region that gives a major contribution to the anomalous Hall effect. 

Actually in the simplest case with the inversion and time-reversal symmetry, the massless Dirac fermion structure may appear at high symmetry points in pair. They contain Chern numbers with opposite signs as in the Honeycomb lattice~\cite{DiVincenzoMele84,Haldane88,KaneMele05} for graphenes. Then, the ferromagnetic moment along the $z$ direction, which breaks the time-reversal symmetry, together with the spin-orbit interaction avoids the crossing of band dispersions and introduces a energy gap separating two dispersions. This transfers the Chern numbers among different band indices. In three dimensions, there occurs a transfer of the Chern numbers among different values of the momentum component $p_z$. In these cases with avoided-crossing of band dispersions, a complete interference among the topological contribution to the Hall conductivity does not occur in general. 

The transfer of the Chern numbers can also occur along momentum curves, for instance, along the $z$ direction parallel to the magnetization. In this case, the two-dimensional massive Dirac-fermion structures studied in the present paper appear at each $p_z$, and $\Delta_0$ continuously changes as a function of $p_z$ with or without a sign change. It is likely that there is at least one such structure across the Fermi level, which is not accompanied by the sign change below the Fermi level. Then, even the integration of the two-dimensional anomalous Hall conductivity over $p_z$ does not lead to a cancellation of the topological contribution, and hence it remains of the order of $e^2/ha$ even for complicated band structures found in first-principles calcualtions~\cite{Fang03,Yao04,Vanderbilt06,YaoFang07,Vanderbilt07}. 

\section{Quantum transport theory for multi-band systems}\label{sec:theory}

We employ the nonequilibrium Green's function method based on the Keldysh formalism~\cite{Mahan,RammerSmith86}, which has recently been reformulated in the gauge-covariant Wigner representation for generic multi-component systems~\cite{Onoda06_ptp}. In the linear order in the electromagnetic field, it clarifies a systematic way of diagrammatically treating the Smr\u{c}ka-St\u{r}eda formula~\cite{SmrckaStreda77,Streda82} and thus the Kubo formula~\cite{Kubo57} with the self-energy and the vertex corrections. Imposing the self-consistency among the Green's function and the self-energy, this automatically satisfies the Ward-Takahashi identity. This formalism also reveals that there appear two mathematically independent self-consistent equations for the linear deviation of quantum distribution function in the electric field, one for the Fermi-surface contribution and the other for the quantum contribution. This is usually not easy to recognize in the integral equation for the vertex correction in the Kubo formalism.

\subsection{Quantum transport theory based on the Keldysh formalism in the gauge-covariant Wigner space}
\label{subsec:overview}

Following our previous paper~\cite{Onoda06_ptp}, we consider the Green's functions $\hat{G}^\alpha$ and the self-energies $\hat{\Sigma}^\alpha$ under the \textit{constant} applied electric field $\bm{E}$, which is taken along the $y$ direction in this paper, i.e., $\bm{E}=(0,E_y)$. Here, the superscripts $\alpha=R$, $A$ and $<$ correspond to the retarded, the advanced and the lesser components, respectively. The Green's functions and the self-energies are considered in the gauge-covariant Wigner space~\cite{Onoda06_ptp} composed of the center-of-mass time ($T$) and space ($\bm{X}$) coordinates and the mechanical energy $\varepsilon$ and momentum $\bm{p}$, which can be obtained by the Fourier transform of the gauge-covariant derivative.
 This gauge-covariant Wigner representation is advantageous over the other choices, since it minimally reduces the arguments of the Green's functions and the self-energies. Namely, in the uniform steady state, it allows the dependence on $X^\mu=(T,\bm{X})$ or $X_\mu=(-T,\bm{X})$ only through the electromagnetic potential $A^\mu(X)=(\phi(\bm{X}),\bm{A}(X))$ or $A_\mu(X)=(-\phi(\bm{X}),\bm{A}(X))$, which can be totally absorbed into the mechanical energy-momentum $p^\mu=(\varepsilon,\bm{p})$ or $p_\mu=(-\varepsilon,\bm{p})$.

Then, the Dyson equations are modified with the applied field as
\begin{subequations}
  \begin{eqnarray}
    \left[\varepsilon\underline{\hat{I}}-\underline{\hat{H}}_0(\bm{p})-\underline{\hat{\Sigma}}(\varepsilon)\right]\star\underline{\hat{G}}(\varepsilon,\bm{p})&=&\underline{\hat{I}},
      \label{eq:_Dyson_:1}\\
      \underline{\hat{G}}(\varepsilon,\bm{p})\star\left[\varepsilon\underline{\hat{I}}-\underline{\hat{H}}_0(\bm{p})-\underline{\hat{\Sigma}}(\varepsilon)\right]&=&\underline{\hat{I}}.
      \label{eq:_Dyson_:2}
  \end{eqnarray}
  \label{eq:_Dyson_}
\end{subequations}
Henceforth, matrices in the Keldysh space are underlined, while those in the band indices are denoted with the hat $\hat{\ }$;
\begin{subequations}
  \begin{eqnarray}
    \underline{\hat{G}}&\equiv&\left(\begin{array}{cc}
      \hat{G}^R & 2\hat{G}^<\\
      0         &  \hat{G}^A
    \end{array}\right),
    \label{eq:_G_}
    \\
    \underline{\hat{\Sigma}}&\equiv&\left(\begin{array}{cc}
      \hat{\Sigma}^R & 2\hat{\Sigma}^<\\
      0         &  \hat{\Sigma}^A
    \end{array}\right),
    \label{eq:_Sigma_}
    \\
    \underline{\hat{H}}_0&\equiv&\left(\begin{array}{cc}
      \hat{H}_0 & 0 \\
      0         &  \hat{H}_0
    \end{array}\right),
    \label{eq:_H_}
    \\
    \underline{\hat{I}}&=&\left(\begin{array}{cc}
      \hat{\sigma}^0 & 0 \\
      0 & \hat{\sigma}^0
    \end{array}\right).
    \label{eq:_I_}
  \end{eqnarray}
\end{subequations}
The symbol $\star$ is the Moyal product of the form,
\begin{equation}
  \star\equiv\exp\left[\frac{i(-e)\hbar}{2}F^{\mu\nu}\left(\overleftarrow{\partial}_{p^\mu}\overrightarrow{\partial}_{p^\nu}-\overleftarrow{\partial}_{p^\nu}\overrightarrow{\partial}_{p^\mu}\right)\right]
  \label{eq:Moyal}
\end{equation}
with the differential operators $\overleftarrow{\partial}$ and $\overrightarrow{\partial}$ operating on the left-hand and the right-hand sides, respectively, and the electromagnetic field tensor $F^{\mu\nu}=\partial_{X_\mu}A^\nu(X)-\partial_{X_\nu}A^\mu(X)$ which is assumed to be constant.
The lesser Green's function and self-energy play roles of the quantum distribution function and the vertex correction, respectively. Since we have assumed that impurity potential has a $\delta$-functional form, the self-energies are local. The distribution function is now fully quantum-mechanical with the $\varepsilon$ dependence, in sharp contrast to the classical Boltzmann transport theory where only its integration over $\varepsilon$ is considered.

$\hat{G}^\alpha(\varepsilon,\bm{p})$ and $\hat{\Sigma}^\alpha(\varepsilon)$ can be expanded in $E_y$ as
\begin{subequations}
  \begin{eqnarray}
    \hat{G}^\alpha(\varepsilon,\bm{p}) &=& \hat{G}^\alpha_0(\varepsilon,\bm{p}) 
    + e\hbar E_y \hat{G}^\alpha_{E_y}(\varepsilon,\bm{p})+O(E_y^2),\ \ \ \ \ 
    \label{eq:g}\\
    \hat{\Sigma}^\alpha(\varepsilon) &=& \hat{\Sigma}^\alpha_0(\varepsilon) 
    + e\hbar E_y \hat{\Sigma}^\alpha_{E_y}(\varepsilon)+O(E_y^2).
    \label{eq:Sigma}
  \end{eqnarray}
\end{subequations}
Henceforth, functionals with the subscripts 0 and $E_y$ denote those in the absence of and the gauge-covariant linear response to $E_y$, respectively. Note that even with the subscript 0, functionals contain the self-energy originating from the impurity scattering. 
$\hat{G}^{R,A}_0$ satisfies the well-known Dyson equation in the absence of the electric field,
\begin{equation}
  \hat{G}^{R,A}_0(\varepsilon,\bm{p})=[\varepsilon-\hat{H}_0(\bm{p})
-\hat{\Sigma}^{R,A}_0(\varepsilon)]^{-1}.
  \label{eq:G^R,A:0}
\end{equation}
The self-consistent equations for $\hat{G}^{R,A,<}_{E_y}$ are obtained by expanding the Moyal product Eq.~(\ref{eq:Moyal}) in the Dyson equation~(\ref{eq:_Dyson_}) in terms of $E_y$~\cite{Onoda06_ptp}. 
It is convenient to decompose $\hat{G}^<_{E_y}$ and $\hat{\Sigma}^<_{E_y}$ 
into two;
\begin{eqnarray}
  \hat{G}^<_{E_y}(\varepsilon,\bm{p})&=&
  \hat{G}^<_{E_y,I}(\varepsilon,\bm{p})\partial_\varepsilon 
  f(\varepsilon)+\hat{G}^<_{E_y,II}(\varepsilon,\bm{p})f(\varepsilon),
  \ \ \ 
  \label{eq:G^<:E}\\
  \hat{\Sigma}^<_{E_y}(\varepsilon)&=&\hat{\Sigma}^<_{E_y,I}(\varepsilon)
  \partial_\varepsilon f(\varepsilon)+\hat{\Sigma}^<_{E_y,II}(\varepsilon)
  f(\varepsilon),
  \label{eq:Sigma^<:E}\\
  \hat{G}^<_{E_y,II}(\varepsilon,\bm{p})&=&\hat{G}^A_{E_y}(\varepsilon,\bm{p})
  -\hat{G}^R_{E_y}(\varepsilon,\bm{p}),
  \label{eq:G^<:E,II}\\
  \hat{\Sigma}^<_{E_y,II}(\varepsilon)&=&\hat{\Sigma}^A_{E_y}(\varepsilon)
  -\hat{\Sigma}^R_{E_y}(\varepsilon).
  \label{eq:Sigma^<:E,II}
\end{eqnarray}
$\hat{G}^<_{E_y,I}$ and $\hat{\Sigma}^<_{E_y,I}$ can be self-consistently 
determined from the quantum Boltzmann equation in the first order in $E_y$,
\begin{eqnarray}
  \lefteqn{\left[\hat{G}^<_{E_y,I},\hat{H}_0\right]+\hat{G}^<_{E_y,I}
\hat{\Sigma}^A_0-\hat{\Sigma}^R_0\hat{G}^<_{E_y,I}}
  \nonumber\\
  &&=\hat{\Sigma}^<_{E_y,I}\hat{G}^A_0-\hat{G}^R_0\hat{\Sigma}^<_{E_y,I}
  -\frac{i}{2}\left[\hat{v}_y,\hat{G}^A_0-\hat{G}^R_0\right]_+
  \nonumber\\
  &&+\frac{i}{2}\left(
  (\hat{\Sigma}^A_0-\hat{\Sigma}^R_0)(\partial_{p_y}\hat{G}^A_0)
  +(\partial_{p_y}\hat{G}^R_0)(\hat{\Sigma}^A_0-\hat{\Sigma}^R_0)\right),
  \ \ \ \ \ 
  \label{eq:G^<:E,I}
\end{eqnarray}
or equivalently,
\begin{eqnarray}
  \hat{G}^<_{E_y,I}&=&\hat{G}_0^R\left[\hat{\Sigma}^<_{E_y,I}-i\partial_{p_y}\left(\hat{H}_0+\frac{1}{2}\left(\hat{\Sigma}^R_0+\hat{\Sigma}_0^A\right)\right)\right]\hat{G}_0^A
  \nonumber\\
  &&+\frac{i}{2}\partial_{p_y}\left(\hat{G}_0^R+\hat{G}_0^A\right),
  \label{eq:G^<:E,I2}
\end{eqnarray}
with the velocity $\hat{v}_i(\bm{p})=\partial_{p_i}\hat{H}_0(\bm{p})$, while 
$\hat{G}^{R,A}_{E_y}$ and $\hat{\Sigma}^{R,A}_{E_y}$ are determined from 
the other self-consistent equation,
\begin{eqnarray}
  \lefteqn{\hat{G}^{R,A}_{E_y}
  =\hat{G}^{R,A}_0\hat{\Sigma}^{R,A}_{E_y}\hat{G}^{R,A}_0}
  \nonumber\\
  &&{}-\frac{i}{2}
  \left(\hat{G}^{R,A}_0\hat{v}_y(\partial_\varepsilon\hat{G}^{R,A}_0)
-(\partial_\varepsilon\hat{G}^{R,A}_0)\hat{v}_y\hat{G}^{R,A}_0\right).
  \label{eq:G^R,A:E}
\end{eqnarray}
Note that the retarded and advanced Green's functions are also modified by the electric field, in contrast to the single-band case. We also stress that the $E_y$-linear deviation of the lesser component of the self-energy, $\hat{\Sigma}^{R,A,<}_{E_y}$, yields the vertex correction and modifies the $\bm{p}$-independent current vertex~\cite{Onoda06_ptp}.

\begin{figure*}[htb]
  \begin{center}
    \includegraphics[width=15.0cm]{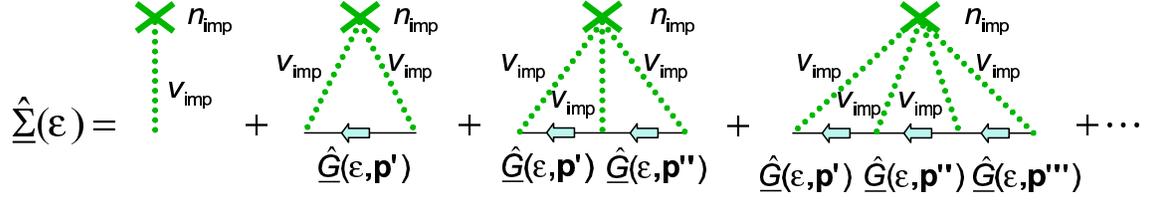}
    \end{center}
  \caption{\label{fig:diagrams}(Color online) Diagrammatic representation of the self-energy in the present self-consistent $T$-matrix approximation in the Keldysh space, which is composed of the infinite series of multiple Born scattering amplitudes. Here, underlined variables $\underline{\hat{G}}$ and $\underline{\hat{\Sigma}}$ are matrices in the Keldysh space. Note that the lesser component gives the integral equation for the vertex correction in the context of the Kubo formalism.}
\end{figure*}
We proceed to the calculation of the self-energies. The side-jump and the skew-scattering contributions can appear from the first and the second Born amplitudes in the vertex correction, respectively~\cite{Bruno01}. Furthermore, for a large impurity potential strength, impurity states may be produced. These nontrivial effects can be fully taken into account up to the linear order in the impurity concentration $n_{\text{imp}}$ by means of the $T$-matrix approximation. Furthermore, the self-consistent calculation of the Green's functions and the self-energies modifies the result and allows for producing a nonperturbative effect of impurity scattering events, as we will see later. The self-consistent $T$-matrix approximation, which is represented by the Feynman diagrams in Fig.~\ref{fig:diagrams}, gives 
\begin{eqnarray}
  \hat{\Sigma}^{R,A}_0(\varepsilon)&=&n_{\rm{imp}}\hat{T}^{R,A}_0(\varepsilon)
  \label{eq:Sigma^R,A:0}\\
  \hat{T}^{R,A}_0(\varepsilon)&=&u_{\text{imp}}\left(1-u_{\text{imp}}\int
\frac{d^2\bm{p}}{(2\pi\hbar)^2}\hat{G}^{R,A}_0(\varepsilon,\bm{p})\right)^{-1}
  \label{eq:g^R,A:0}
\end{eqnarray}
for the zeroth-order in $E_y$ and
\begin{eqnarray}
  \hat{\Sigma}^<_{E_y,I}(\varepsilon)\!\!&=&\!\!n_{\rm{imp}}
\hat{T}^R_0(\varepsilon)\!\!\int\!\!\frac{d^2\bm{p}}{(2\pi\hbar)^2}
\hat{G}^<_{E_y,I}(\varepsilon,\bm{p})\hat{T}^A_0(\varepsilon)
  \label{eq:Sigma^<:E,I}\\
  \hat{\Sigma}^{R,A}_{E_y}(\varepsilon)\!\!&=&\!\!n_{\rm imp}
\hat{T}^{R,A}_0(\varepsilon)\!\!\int\!\!\frac{d^2\bm{p}}{(2\pi\hbar)^2}
\hat{G}^{R,A}_{E_y}(\varepsilon,\bm{p})\hat{T}^{R,A}_0(\varepsilon)
  \label{eq:Sigma^R,A:E}
\end{eqnarray}
for the first-order in $E_y$.

\begin{figure}[htb]
  \begin{center}
    \includegraphics[width=7.2cm]{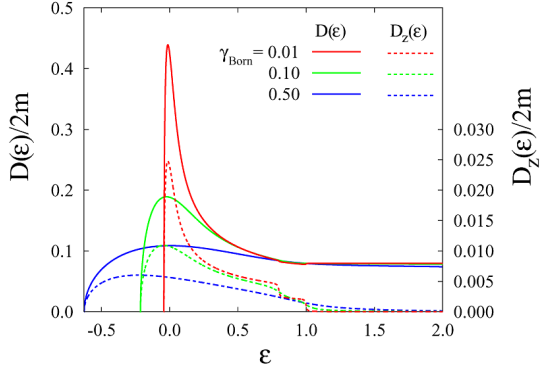}
  \end{center}
  \caption{(Color online) The results of the local electron and spin density of states, $D(\varepsilon)$ and $D_z(\varepsilon)$, respectively, in the self-consistent $T$-matrix approximation for $v=3.59$, $\Delta_0=0.1$, and $2mu_{\text{imp}}=0.2$.}
    \label{fig:DOS}
\end{figure}

\subsection{Equilibrium properties}
\label{subsec:theory:0}

We first solve Eqs.~(\ref{eq:G^R,A:0}), (\ref{eq:Sigma^R,A:0}) and (\ref{eq:g^R,A:0}) for the self-consistent $T$-matrix approximation to obtain the equilibrium Green's functions $\hat{G}^{R,A}_0$ and self-energies $\hat{\Sigma}^{R,A}_0$. All the momentum integrations are performed analytically for each value of the energy $\varepsilon$ as given in Appendix~\ref{app:G^R_0}, and then the numerical iteration is repeated until the convergence is reached.
Figure~\ref{fig:DOS} shows thus obtained local electron and spin density of states,
\begin{eqnarray}
  D(\varepsilon)&\equiv& -\frac{1}{\pi}\int\frac{d^2\bm{p}}{(2\pi\hbar)^2}\Im\,\text{Tr}\left[\hat{G}_0^R(\varepsilon,\bm{p})\right],
  \label{eq:DOS}\\
  D_z(\varepsilon)&\equiv& -\frac{1}{\pi}\int\frac{d^2\bm{p}}{(2\pi\hbar)^2}\Im\,\text{Tr}\left[\hat{G}_0^R(\varepsilon,\bm{p})\hat{\sigma}^z\right],
  \label{eq:DOSz}
\end{eqnarray}
with a set of parameters $v=3.59$, $\Delta_0=0.1$, and $2mu_{\text{imp}}=0.2$ for three choices of $\gamma_{\text{Born}}=\hbar/\tau_{\text{Born}}\equiv mn_{\text{imp}}u_{\text{imp}}/\hbar^2=0.01$, 0.10, and 0.50. Here, $\gamma_{\text{Born}}$ represents the first Born scattering amplitude in the case of $v=\Delta_0=0$.
The increase of the impurity potential strength $u_{\text{imp}}$ lowers the bottom of the band due to the broadening of the electron spectral functions and also smears out the singularity which is originally present at the energy levels of the majority and the minority bands at $\bm{p}=0$, i.e., $\varepsilon=\varepsilon_\pm(p=0)$, without the impurities. This broadened spectral feature is beyond the semi-classical approximation where the electron spectral function has a $\delta$-functional form. This also plays a crucial role in eliminating an unphysical singularity in the thermoelectric Hall conductivity $\alpha_{xy}$ at $E_F=\varepsilon_\pm(p=0)$ and the discontinuity in $\sigma_{xy}$ at $E_F=\varepsilon_-(p=0)$, as shown later.

\subsection{Field-induced change of quantum distribution function, and the electric conductivity tensor}
\label{subsec:theory:1}

Next, the self-consistent results for $\hat{G}^R_0(\varepsilon,\bm{p})$ and $\hat{\Sigma}^R_0(\varepsilon)$ are plugged into Eqs.~(\ref{eq:G^<:E,I}) and (\ref{eq:Sigma^<:E,I}) to calculate $\hat{G}^<_{E_y,I}$ and $\hat{\Sigma}^<_{E_y,I}$ self-consistently, and into Eqs.~(\ref{eq:G^R,A:E}) and (\ref{eq:Sigma^R,A:E}) to obtain the self-consistent solution for $\hat{G}^{R,A}_{E_y}$ and $\hat{\Sigma}^{R,A}_{E_y}$. Details of the calculations of $\hat{G}^<_{E_y,I}$ and $\hat{\Sigma}^<_{E_y,I}$, and $\hat{G}^{R,A}_{E_y}$ and $\hat{\Sigma}^{R,A}_{E_y}$ are given in Appendices~\ref{app:sigma^I} and \ref{app:sigma^II}, respectively. Then, $\hat{G}^<_{E_y,II}$ is calculated from $\hat{G}^{R,A}_{E_y}$ via Eq.~(\ref{eq:G^<:E,II}). 

Finally, the conductivity tensor is obtained as
\begin{eqnarray}
  \sigma_{ij}^{\text{Tot}}&=&e^2\hbar\int\!\frac{d\varepsilon}{2\pi i}\int\!\frac{d^2\bm{p}}{(2\pi\hbar)^2}\text{Tr}\left[\hat{v}_i(\bm{p})\hat{G}_{E_j}^<(\varepsilon,\bm{p})\right]
  \nonumber\\
  &=&\sigma_{ij}^I+\sigma_{ij}^{II}
  \label{eq:sigma}
\end{eqnarray}
with $i,j=x,y$, where $\sigma_{ij}^{\text{Tot}}$ has been decomposed into the Fermi-surface contribution $\sigma_{xy}^I$ and the quantum contribution $\sigma_{xy}^{II}$
\begin{eqnarray}
  \sigma_{ij}^I &=& e^2\hbar\int\!\frac{d\varepsilon}{2\pi i}\int\!\frac{d\bm{p}}{(2\pi\hbar)^2}
  \text{Tr}\left[\hat{v}_i(\bm{p})\hat{G}^<_{E_j,I}(\varepsilon,\bm{p})\right]
  \partial_\varepsilon f(\varepsilon)
  \nonumber\\
  &\to& -\frac{e^2\hbar}{2\pi i}
  \int\!\!\frac{d^2\bm{p}}{(2\pi\hbar)^2}{\rm Tr}\!
  \left[\hat{v}_i(\bm{p}) \hat{G}^<_{E_j,I}(\mu,\bm{p})\right],
  \label{eq:sigma^I}\\
  \sigma_{ij}^{II} &=& e^2\hbar\int\!\!\frac{d\varepsilon}{2\pi i}
  \int\!\!\frac{d^2\bm{p}}{(2\pi\hbar)^2}{\rm Tr}\!
  \left[\hat{v}_i(\bm{p}) \hat{G}^<_{E_j,II}(\varepsilon,\bm{p})\right] f(\varepsilon)
  \nonumber\\
  &\to& e^2\hbar\int^\mu_{-\infty}\!\!\!\frac{d\varepsilon}{2\pi i}
  \int\!\!\frac{d^2\bm{p}}{(2\pi\hbar)^2}{\rm Tr}\!
  \left[\hat{v}_i(\bm{p}) \hat{G}^<_{E_j,II}(\varepsilon,\bm{p})\right].
  \nonumber\\
  \label{eq:sigma^II}
\end{eqnarray}
The second lines in Eqs.~(\ref{eq:sigma^I}) and (\ref{eq:sigma^II}) are obtained in the zero temperature limit.
Here, $\sigma_{ij}^I$ is completely determined by the Fermi-surface properties at the zero temperature, while $\sigma_{ij}^{II}$ contains the whole Fermi-sea properties and contributes to only the Hall conductivity.
Now, we can calculate $\sigma_{ij}^I$ and $\sigma_{ij}^{II}$ separately by substituting Eq.~(\ref{eq:G^<:E,I}) into Eq.~(\ref{eq:sigma^I}) and Eqs.~(\ref{eq:G^<:E,II}) and (\ref{eq:G^R,A:E}) into Eq.~(\ref{eq:sigma^II}), respectively
We note that for our practical calculations, all the momentum integrations in Eqs.~(\ref{eq:sigma^I}) and (\ref{eq:sigma^II}) are performed analytically, and then the remaining energy integration is performed numerically in Eq.~(\ref{eq:sigma^II}). Details of the calculations are given in Appendices~\ref{app:sigma^I} and \ref{app:sigma^II}.

Apart from the mathematical separation as given in Eq.~(\ref{eq:sigma}), it is useful to introduce another decomposition scheme to distinguish the mechanisms. Effects of the scattering events result in the equilibrium self-energy $\hat{\Sigma}^{R,A}_0$ and the vertex corrections associated with $\hat{\Sigma}^{R,A,<}_{E_y}$. The extrinsic contribution can be ascribed to the vertex corrections, while the equilibrium self-energy correction yields the modification of the equilibrium electronic structure like the quasiparticle dispersion and the damping rate. The latter only modifies the intrinsic contribution from that obtained for the perfect crystal. Then, it is meaningful to separate the total conductivity into the intrinsic and extrinsic parts. In particular, the intrinsic part is defined as the contribution that survives without the vertex correction,
\begin{equation}
  \sigma_{xy}^{\text{Int}}=\sigma_{xy}^{I\ \text{Int}}+\sigma_{xy}^{II\ \text{Int}},
  \label{eq:sigma_xy^Int}
\end{equation}
where 
\begin{eqnarray}
  \lefteqn{\sigma_{xy}^{I\ \text{Int}}
  =-\frac{e^2\hbar}{2}\int\!\frac{d\varepsilon}{2\pi}\partial_\varepsilon f(\varepsilon)
  \int\!\frac{d\bm{p}}{(2\pi\hbar)^2}\ }
  \nonumber\\
  &&\times\text{Tr}
  \Bigl[\hat{v}_i(\bm{p})\hat{G}^R_0(\varepsilon,\bm{p})\hat{v}_j(\bm{p})\left(\hat{G}^A_0(\varepsilon,\bm{p})-\hat{G}^R_0(\varepsilon,\bm{p})\right)
    \nonumber\\
    &&\hspace*{10pt}
    {}-\hat{v}_i(\bm{p})\left(\hat{G}^A_0(\varepsilon,\bm{p})-\hat{G}^R_0(\varepsilon,\bm{p})\right)\hat{v}_j(\bm{p})\hat{G}^A_0(\varepsilon,\bm{p})\Bigr]
  \ \ \ \ \ \ 
  \label{eq:sigma_xy^I:int1}
\end{eqnarray}
and
\begin{eqnarray}
  \sigma^{II\ \text{Int}}_{xy} 
  &=&e^2\hbar\int\!\frac{d\varepsilon}{2\pi}f(\varepsilon)\int\!\frac{d\bm{p}}{(2\pi\hbar)^2}
  \ 
  \nonumber\\
  &&\times\text{Tr}\left[\hat{v}_x(\bm{p})\hat{G}^A_0(\varepsilon,\bm{p})\hat{v}_y(\bm{p})(\partial_\varepsilon\hat{G}^A_0(\varepsilon,\bm{p}))
    \right.\nonumber\\
    &&\hspace*{12pt}{}-\hat{v}_x(\bm{p})(\partial_\varepsilon\hat{G}^A_0(\varepsilon,\bm{p}))\hat{v}_y(\bm{p})\hat{G}^A_0(\varepsilon,\bm{p})
    \nonumber\\
    &&\hspace*{12pt}{}-\hat{v}_x(\bm{p})\hat{G}^R_0(\varepsilon,\bm{p})\hat{v}_y(\bm{p})(\partial_\varepsilon\hat{G}^R_0(\varepsilon,\bm{p}))
    \nonumber\\
    &&\left.\hspace*{11pt}{}+\hat{v}_x(\bm{p})(\partial_\varepsilon\hat{G}^R_0(\varepsilon,\bm{p}))\hat{v}_y(\bm{p})\hat{G}^R_0(\varepsilon,\bm{p})\right]
  \ \ \
  \label{eq:sigma_xy^II:int}
\end{eqnarray}
are obtained from Eqs.~(\ref{eq:sigma^I}) and (\ref{eq:sigma^II}) by ignoring $\hat{\Sigma}^<_{I,E_y}$ and $\hat{\Sigma}^{R,A}_{E_y}$ in Eqs.~(\ref{eq:G^<:E,I2}) and (\ref{eq:G^R,A:E}), respectively. 
The extrinsic contribution is then calculated as the difference 
\begin{equation}
  \sigma_{xy}^{\text{Ext}}=\sigma_{xy}^{\text{Tot}}-\sigma_{xy}^{\text{Int}}.
  \label{eq:sigma_xy^Ext}
\end{equation}

Especially, when the relaxation rate vanishes, $\sigma_{ij}^{I\ \text{Int}}$ is analytically expressed as~\cite{Onoda06_prl}
\begin{eqnarray}
  \lefteqn{\sigma^{I\ \text{Int}}_{ij}(\tau\to\infty)
  =-\epsilon_{ij\ell}\frac{e^2\hbar}{2}\int\!\frac{d\bm{p}}{(2\pi\hbar)^2}\sum_{n,n'}(\varepsilon_n(\bm{p})-\varepsilon_{n'}(\bm{p}))}
  \nonumber\\
  &&\times\partial_\varepsilon f(\varepsilon_n(\bm{p})){\rm Im}\left(\langle n\bm{p}|\bm{\nabla}_p|n'\bm{p}\rangle\times\langle n'\bm{p}|\bm{\nabla}_p|n\bm{p}\rangle\right)_\ell,\ \ \ \ \ 
  \label{eq:sigma_xy^I:int2}
\end{eqnarray}
which can be directly derived from Eq.~(\ref{eq:G^<:E,I}) or (\ref{eq:G^<:E,I2}).
This and $\sigma_{xy}^{II\ \text{Int}}(\tau\to\infty)$ compose the Berry-curvature contribution, i.e., the TKNN formula given by Eq.~(\ref{eq:TKNN}). Namely, provided that there is no singular energy dependence in the self-energy $\hat{\Sigma}_0^{R,A}$, the relation
\begin{equation}
  \sigma_{xy}^{\text{TKNN}}=\sigma_{xy}^{I\ \text{Int}}(\tau\to\infty)+\sigma_{xy}^{II\ \text{Int}}(\tau\to\infty)
  \label{eq:sigma_xy^II}
\end{equation}
holds, as addressed previously~\cite{Onoda06_prl,Sinova}. Then, the integration by parts shows that $\sigma_{xy}^{\text{Int}}$ can be related to the Fermi-surface properties~\cite{Haldane04}. 

In fact, even with an infinitesimally small impurity concentration, i.e., in the clean limit ($n_{\rm imp}\to0$), $\sigma_{xy}^{II}$ may be suppressed from the TKNN result Eq.~(\ref{eq:TKNN}) by the contribution from the bottom of the bands due to a nearly singular energy dependence of the self-energy $\hat{\Sigma}^{R,A}_0$, as is shown mathematically in Appendix~\ref{app:G^R_0} and is plotted in Fig.~\ref{fig:DOS}. This reflects that the momentum $\bm{p}$ and thus the Berry curvature are no longer good quantum numbers in the presence of the impurity potential. 
In terms of a constant relaxation-rate approximation, $\sigma_{xy}^{\text{Int}}$ can be calculated from the first principles~\cite{Fang03,Yao04,Vanderbilt06,YaoFang07,Vanderbilt07}. Even with a nontrivial energy-dependent self-energy, the first-principles calculation is possible, for instance, using the GW approximation~\cite{Ferdi}.

Physically, the intrinsic contribution actually corresponds to $\sigma_{xy}(\omega)$ in the limit where $\tau\to\infty$ and subsequently $\omega\to0$. In the metallic case, it disagrees with the result in the real DC limit ($\omega\to0$ then $\tau\to\infty$) which is directly relevant to the transport properties. When the Fermi level is located within the energy gap, $\sigma_{xy}^I$ vanishes and $\sigma_{xy}^{II}$ agrees with Eq.~(\ref{eq:TKNN}). In general, $\sigma_{xy}^{II}$ is robust against the scattering and thus the vertex corrections, and hence we can regard $\sigma_{xy}^{II}$ as an intrinsic contribution even in the presence of impurities. Actually, for the present model given by Eq.~(\ref{eq:Hs}), the vertex correction to $\sigma_{xy}^{II}$, namely, the effect of $\hat{\Sigma}^<_{E_j,II}$ on $\sigma_{xy}^{II}$, is canceled out, as shown in Appendix~\ref{app:sigma^II}. However, the Fermi-surface contribution $\sigma_{xy}^I$ is strongly affected by a disspation originating from the vertex correction in the clean limit, and the extrinsic contribution plays a crucial role. 

The expression given by Eq.~(\ref{eq:sigma}) together with Eqs.~(\ref{eq:sigma_xy^I:int1}) and (\ref{eq:sigma_xy^II:int}) coincide with the Sm\u{r}cka-St\u{r}eda formula~\cite{SmrckaStreda77,Streda82}. Moreover, it is remarkable that in the presence of scattering, this approach based on Eqs.~(\ref{eq:sigma}), (\ref{eq:sigma^I}), and (\ref{eq:sigma^II}) provides the diagrammatic treatment for the Sm\u{r}ka-St\u{r}eda formula~\cite{SmrckaStreda77,Streda82}, as previously noted~\cite{Onoda06_prl,Onoda06_ptp}. Here, instead of diagonalizing the impurity Hamiltonian and expressing the conductivity tensor in the diagonalized basis, we have taken into account the self-energy and the vertex corrections due to the impurities, which correspond to $\hat{\Sigma}_0^{R,A}$ and $\hat{\Sigma}_{E_y}^<$, respectively. 

It is also important that formally there exist two independent self-consistent equations for the total quantum distribution function $\hat{G}^<_{E_y}(\varepsilon,\bm{p})$: one is for $\hat{G}^{<}_{E_y,I}(\varepsilon,\bm{p})$ and the other for $\hat{G}^{R,A}_{E_y}(\varepsilon,\bm{p})$, in addition to the equilibrium Green's function $\hat{G}^<_0(\varepsilon,\bm{p})=(\hat{G}^A_0(\varepsilon,\bm{p})-\hat{G}^R_0(\varepsilon,\bm{p}))f(\varepsilon)$. Therefore, one can not correctly obtain $\hat{G}^<_{E_y}$ without solving all these self-consistent equations, in general.

\subsection{Mott rule and Wiedemann-Franz law}
\label{subsec:theory:Mott,Wiedemann-Franz}

Let us consider the electric current $\bm{J}$ and thermal current $\bm{J}^q$ due to the electric field $\bm{E}$ and the temperature gradient $\bm{\nabla}T$ applied to the sample. Up to the linear order in $\bm{E}$ and $\bm{\nabla}T$, they are conventionally written as
\begin{subequations}
\begin{eqnarray}
  J_i &=& \sum_j\left(\sigma_{ij}E_j+\alpha_{ij}(-\partial_j T)\right),
  \label{eq:J}\\
  J^q_i &=& \sum_j\left(T\alpha_{ij}E_j+\kappa_{ij}(-\partial_j T)\right).
\label{eq:J^q}
\end{eqnarray}
\end{subequations}
Here, $\sigma_{ij}$ is nothing but the electric conducticity tensor, and $\alpha_{ij}$ and $\kappa_{ij}$ are the thermoelectric and the thermal conducticity tensors, respectively. The anomalous thermoelectric Hall effect, i.e., the anomalous Nernst-Ettingshausen effect, is characterized by $\alpha_{xy}\ne0$, while the anomalous thermal Hall effect, i.e., the anomalous Luduc-righi effect, by $\kappa_{xy}\ne0$.

Sm\u{r}cka-St\u{r}eda~\cite{SmrckaStreda77} proved that the Mott rule
\begin{subequations}
\begin{equation}
  \alpha_{ij}=\frac{\pi^2k_B^2}{3(-e)}T\left[\frac{d}{d\varepsilon}\sigma_{ij}(\varepsilon)\right]_{\varepsilon=\mu},
  \label{eq:Mott}
\end{equation}
 and the Wiedemann-Franz law
\begin{equation}
    \kappa_{ij}=\frac{\pi^2k_B^2}{3e^2}T\sigma_{ij}(\mu),
    \label{eq:Wiedemann-Franz}
\end{equation}
\end{subequations}
generally hold in the low-temperature limit in the absence of inelastic scattering, with the Bolzmann constant $k_B$, the temperature $T$, and the electric conductivity tensor $\sigma_{ij}(\mu)$ at the chemical potential $\mu$.

The $T$-linear coefficient to $\kappa_{ij}/T$ is just proportional to $\sigma_{ij}$, and does not contain new information. On the other hand, the low-temperature limit of $\alpha_{ij}/T$ is proportional to the derivative of $\sigma_{ij}(\mu)$ with respect to the chemical potential. Therefore, it is calculated from
\begin{equation}
  \alpha_{ij}^{\text{tot}}=\alpha_{ij}^I+\alpha_{ij}^{II}
  \label{eq:alpha}
\end{equation}
with
\begin{eqnarray}
  \frac{\alpha_{ij}^I}{T} &\to& -\frac{\pi^2k_B^2e\hbar}{6\pi i}
  \int\!\!\frac{d^2\bm{p}}{(2\pi\hbar)^2}{\rm Tr}\!
  \left[\hat{v}_i(\bm{p}) \frac{d}{d\mu}\hat{G}^<_{E_j,I}(\mu,\bm{p})\right],
  \nonumber\\
  \label{eq:alpha^I}\\
  \frac{\alpha_{ij}^{II}}{T} &\to& \frac{\pi^2k_B^2e\hbar}{6\pi i}
  \int\!\!\frac{d^2\bm{p}}{(2\pi\hbar)^2}{\rm Tr}\!
  \left[\hat{v}_i(\bm{p}) \hat{G}^<_{E_j,II}(\mu,\bm{p})\right].
  \label{eq:alpha^II}
\end{eqnarray}
Note that it includes the Berry-phase curvature at the Fermi level~\cite{YaoFang07,Haldane04}.

\subsection{Connection with the semi-classical approach}
\label{subsec:semiclassical}

Now it is important to clarify the relation between the present fully quantum-mechanical approach~\cite{Onoda06_prl,Onoda06_ptp} and the semi-classical approaches modified with the Berry phase in the momentum space~\cite{SundaramNiu99,Sinova}.
From the present formalism, if we ignore the self-energy corrections, i.e., $\hat{\Sigma}^{R,A}$, the transport equations~(\ref{eq:G^<:E,I}) and (\ref{eq:G^R,A:E}) can be readily integrated over the energy $\varepsilon$. Note that when we calculate the extrinsic transport current from Eq.~(\ref{eq:G^<:E,I}), we need to multiply the both sides of the equation by the typical relaxation rate $\tau$ to maintain the meaning. Then, they are expressed in terms of the semi-classical distribution function but of the matrix form,
\begin{eqnarray}
  \hat{F}(\bm{p})=\int\!\frac{d\varepsilon}{2\pi i}
  \left[\left(\hat{G}^A_0(\varepsilon,\bm{p})-\hat{G}^R_0(\varepsilon,\bm{p})\right)f(\varepsilon)
    +\hat{G}^<_{E_y}(\varepsilon,\bm{p})\right].
  \nonumber\\
  \label{eq:F}
\end{eqnarray}
The first term, i.e., the equilibrium part, can be diagonalized by the band representation with the band index $n$ and the dispersion $\varepsilon_n(\bm{p})$. Then, the energy integration yields $f(\varepsilon_n(\bm{p}))$. Though this procedure does not diagonalize the second term, i.e., the nonequilibrium part, this semi-classical approach usually serves as a good approximation to the longitudinal conductivity.

Similar techniques have been used to calculate the anomalous Hall conductivity in the same model, using the Kubo formula and the semi-classical Boltzmann theory~\cite{Sinova}. However, whichever method is used, ignoring the self-energy correction in the energy integration of the equilibrium Green's function in the Kubo formalism or using the semi-classical distribution function in the semi-classical Boltzmann transport theory sometimes leads to an unphysical singularity, as we will explain in later sections. Unfortunately, this is the case in the present model, and one needs to include seriously the lifetime broadening of the quasiparticles. In this respect, the present fully quantum-mechanical approach gives a powerful theoretical formalism by which one can directly treat non-trivial quasiparticle spectra modified by the self-energy correction, which always eliminates such singularity in the presence of a finite scattering strength.

\section{Anomalous electric and thermal Hall effects}
\label{sec:AHE}

In this section, we will show the results for the anomalous electric Hall conductivity $\sigma_{xy}$. The low-temperature value of the anomalous thermal Hall conductivity $\kappa_{xy}$ can be directly obtained from $\sigma_{xy}$ with the universal proportionality constant via the Wiedmann-Frannz law given by Eq.~(\ref{eq:Wiedemann-Franz}).

\subsection{Global dependence of extrinsic and intrinsic contributions on the Fermi energy and the scattering amplitude}
\label{subsec:AHE:E_F,tau}

\begin{figure}
  \begin{center}
    \includegraphics[width=8.0cm]{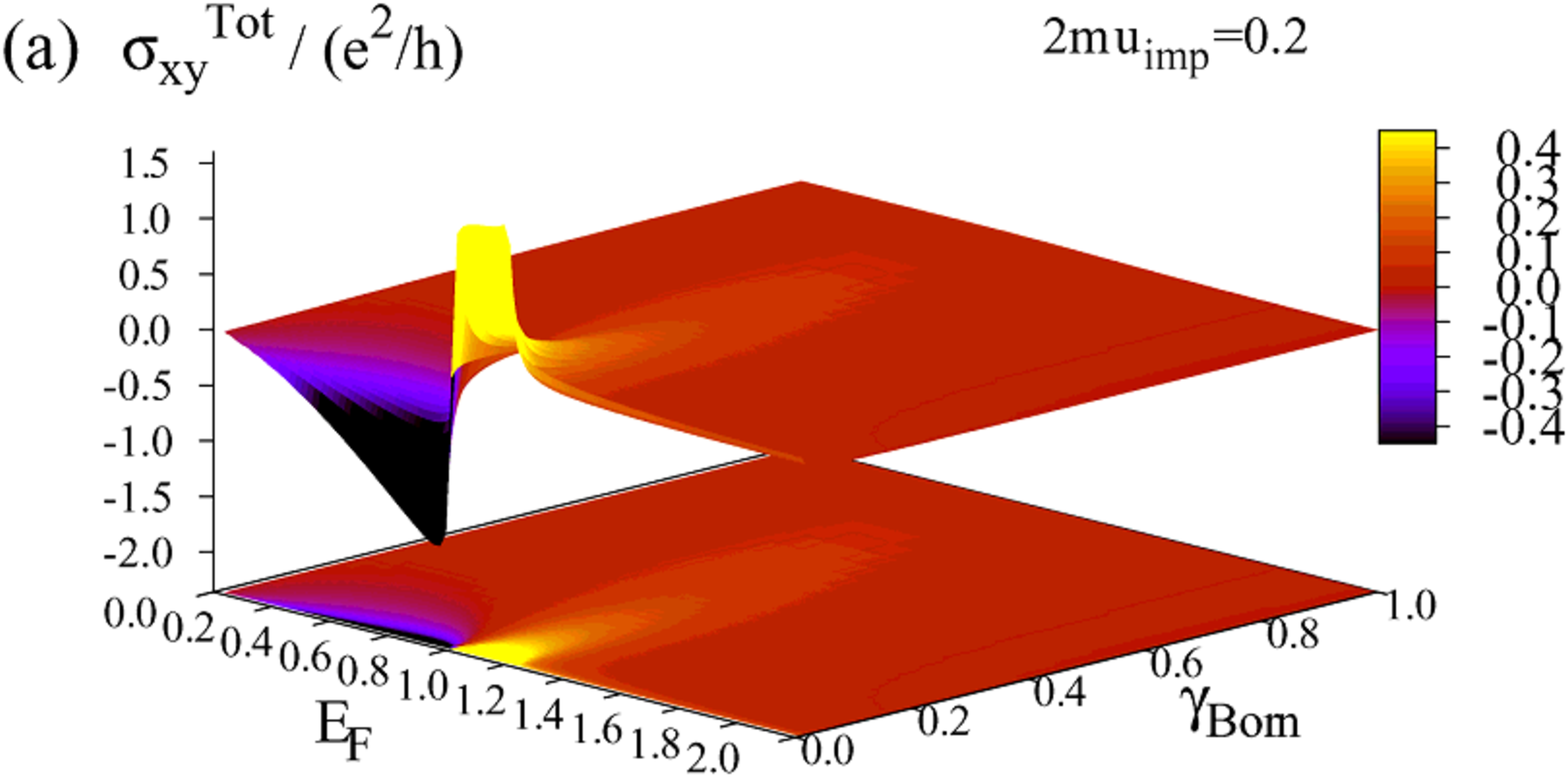}
    \includegraphics[width=8.0cm]{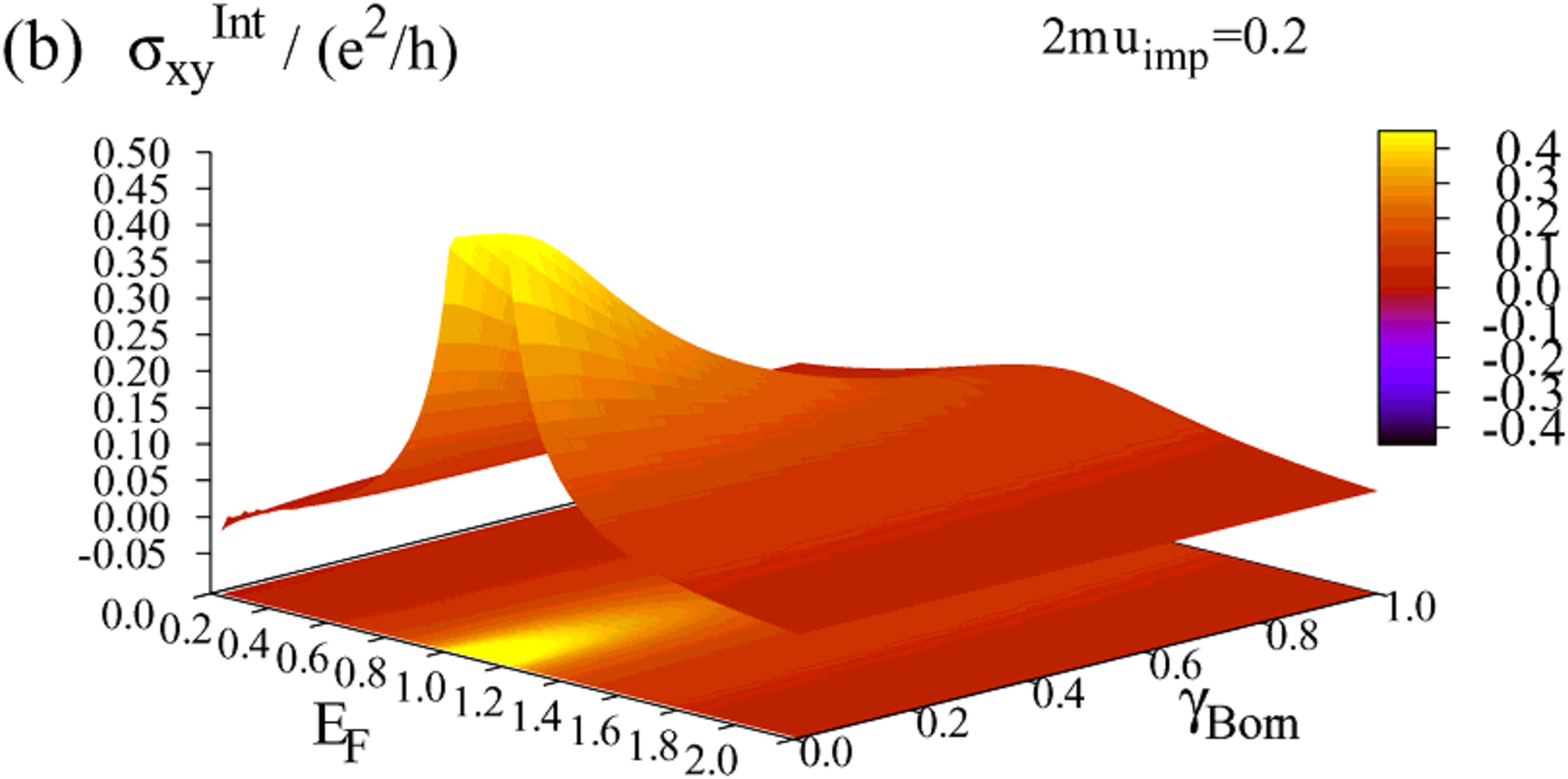}
    \includegraphics[width=8.0cm]{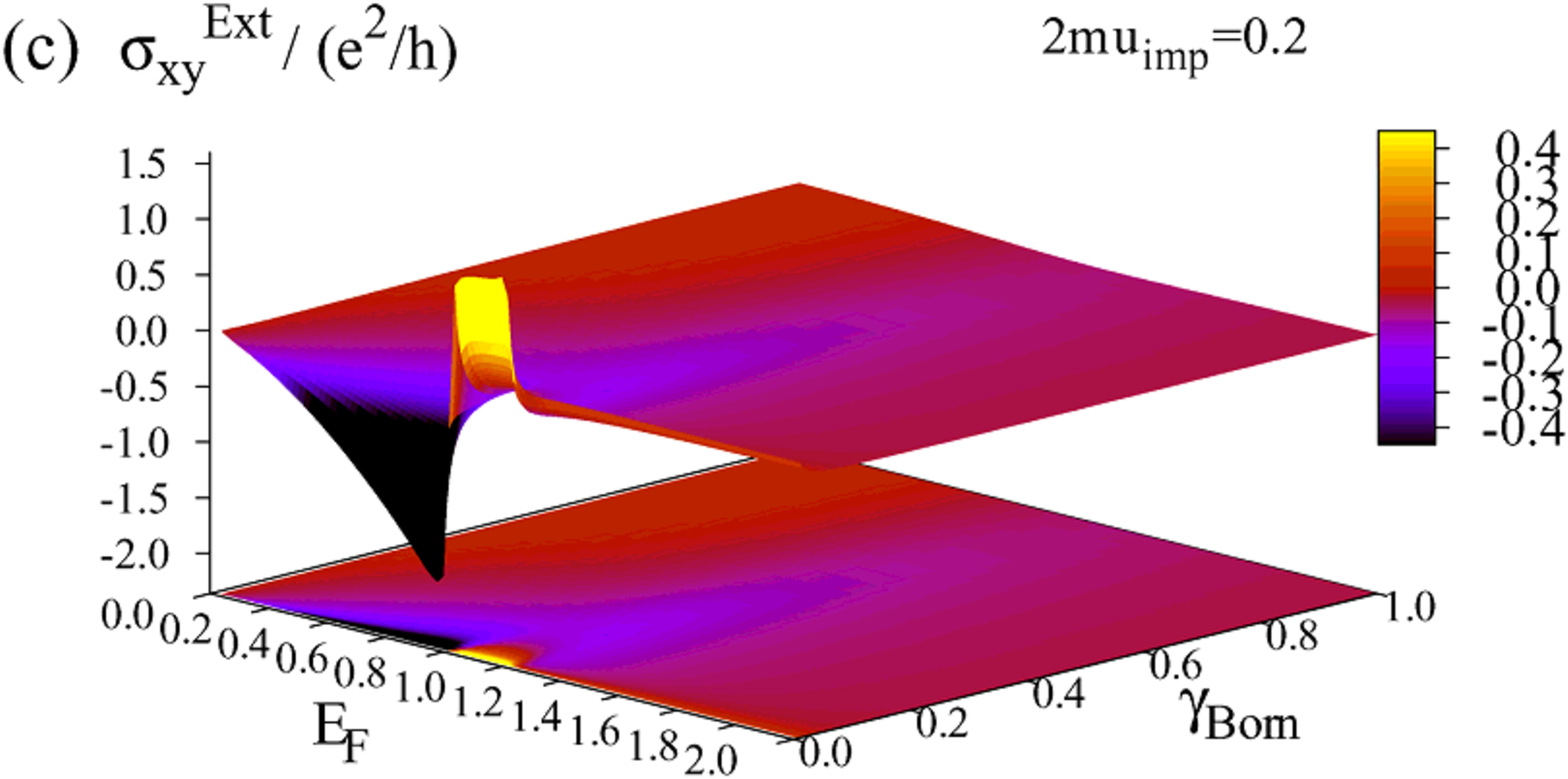}
  \end{center}
  \caption{(Color online) (a) The total anomalous Hall conductivity $\sigma_{xy}^{\text{Tot}}$, (b) the intrisic contribution $\sigma_{xy}^{\text{Int}}$, and (c) the extrinsic contribution $\sigma_{xy}^{\text{ext}}$ as functions of the Fermi energy $E_F$ and the Bron scattering amplitude $\gamma_{\text{Born}}$ in an energy unit of $E_{\rm res}=1.0$. The parameters are choisen as $v=3.59$, $\Delta_0=0.1$, and $2mu_{\text{imp}}=0.2$. Note the difference of the scales for $\sigma_{xy}$ in (a), (b), and (c).}
\label{fig:Hall1}
\end{figure}

Figure~\ref{fig:Hall1} (a) shows the numerical results on the total anomalous Hall conductivity $\sigma_{xy}^{\text{Tot}}=\sigma_{xy}^I+\sigma_{xy}^{II}$ as a function of the Fermi energy $E_F$ and the first Born scattering amplitude $\gamma_{\text{Born}}=\hbar/\tau_{\text{Born}}\equiv n_{\rm imp}u_{\text{imp}}^2 m$ for a typical set of parameters, $\Delta_0=0.1$, $2mu_{\text{imp}}=0.6$, and $2mv^2=3.59$ in an energy unit of $E_{\rm res}=1.0$. Henceforth, the energy cutoff is taken as $E_c=3.0$ and $\gamma_{\text{Born}}$ is varied by changing the impurity concentration $n_{\text{imp}}$ which can be directly controlled in experiments, while the potential strength $u_{\text{imp}}$ is fixed.

In the clean limit $\gamma_{\text{Born}}=\hbar/\tau_{\text{Born}}\to0$, $\sigma_{xy}^{\text{Tot}}$ tends to increase rapidly in accordance with the extrinsic skew-scattering scenario (see Eq.~(\ref{eq:skew})). Strength of the divergence is proportional to $E_F$ in the low electron-density limit, and the sign is inverted around $E_F=\varepsilon_+(0)=E_{\rm res}-2\Delta_0$. Sign of this skew-scattering contribution $\sigma_{xy}^{\text{skew}}$ also changes by that of $u_{\text{imp}}$. It is also evident from Fig.~\ref{fig:Hall1} that $\sigma_{xy}^{\text{skew}}$ is significantly reduced when the both bands are partially occupied, i.e., $\mu>E_{\text{res}}=\varepsilon_-(p=0)$. If the self-energy corrections $\hat{\Sigma}^{R,A}_0$ are approximated by constant relaxation rates, $\sigma_{xy}^{\text{skew}}$ might completely vanish when the both bands are partially occupied and a discontinuity would also appear when the Fermi level crosses the bottom of the minority band~\cite{Sinova}. In fact, the self-energy determined self-consistently with the Green's function causes a small but finite skew-scattering contribution even in this case. This is natural since the Fermi surface is now not strictly defined and the quasiparticles around the Fermi level participate in the extrinsic dissipative Hall current through the asymmetric scattering. We will discuss this issue in detail later in Sec.~\ref{sec:others}.

To identify the intrinsic and extrinsic contributions quantitatively, we adopt the separation scheme for the extrinsic and the intrinsic parts defined via Eqs.~(\ref{eq:sigma_xy^Int}) and (\ref{eq:sigma_xy^Ext}). The intrinsic part $\sigma_{xy}^{\text{Int}}$ is plotted in Fig.~\ref{fig:Hall1} (b) for the same set of parameters. Under the resonant condition for $E_F$ being around the range $[\varepsilon_-(p=0)-2\Delta_0,\varepsilon_-(p=0)]$, $\sigma_{xy}^{\text{Int}}$ becomes of the order of $e^2/2h$. With increasing the scattering amplitude $\gamma_{\text{Born}}$, it only gradually decreases due to the damping of quasiparticles. Off the resonance, $\sigma_{xy}^{\text{Int}}$ is significantly reduced by a small factor $E_{\text{SO}}D$.

On the other hand, the extrinsic part $\sigma_{xy}^{\text{Ext}}$ is shown in Fig.~\ref{fig:Hall1} (c). It is evident that the extrinsic skew-scattering process always yields a dominant contribution to $\sigma_{xy}$ in the clean limit. However, with increasing the relaxation $\gamma_{\text{Born}}$ by the increase of $n_{\text{imp}}$, $\sigma_{xy}^{\text{skew}}$ rapidly decays and becomes comparable to or even smaller than $\sigma_{xy}^{\text{Int}}$, indicating an crossover from the extrinsic skew-scattering regime to the intrinsic regime. This rapid decay reflects that the skew-scattering contribution originates from intra-band processes and hence the skewness factor $S$ remains of the order of $E_{\text{SO}}u_{\text{imp}}D/E_F$.

\begin{figure}
  \begin{center}
    \includegraphics[width=8.0cm]{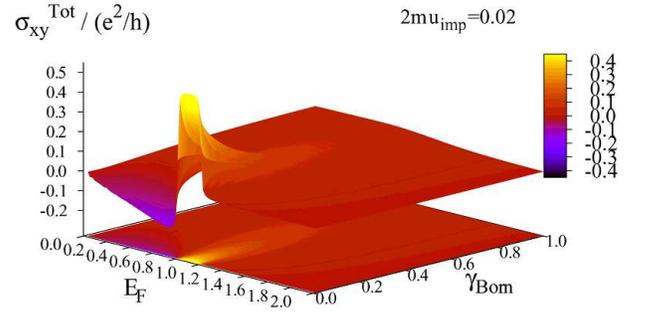}
  \end{center}
  \caption{(Color online) The total anomalous Hall conductivity $\sigma_{xy}^{\text{Tot}}$ as a function of $E_F$ and $\hbar/\tau$ in an energy unit of $E_{\rm res}=1.0$. The parameters are choisen as $v=3.59$, $\Delta_0=0.1$, and $2mu_{\text{imp}} = 0.02$. Note the difference of the scale for $\sigma_{xy}$ compared with Fig.~\ref{fig:Hall1} (a).}
\label{fig:Hall2}
\end{figure}

In Fig.~\ref{fig:Hall2}, we also show the results obtained with the same set of parameters except a smaller value of the impurity potential strength $2mu_{\text{imp}}=0.02$. Compared with the case of a larger $u_{\text{imp}}$ shown in Fig.~\ref{fig:Hall1}, the skew-scattering contribution is smaller, while the intrinsic contribution almost remains the same. In particular, in the level of the (self-consistent) Born approximation instead of the (self-consistent) $T$-matrix approximation, $\sigma_{xy}^{\text{Int}}$ depends on $u_{\text{imp}}$ and $n_{\text{imp}}$ only through $\gamma_{\text{Born}}$. The reduction of the skew-scattering contribution is also natural since as explained below Eq.~(\ref{eq:skew}) in Sec.~\ref{sec:intro}, it is proportional to $1/n_{\text{imp}}u_{\text{imp}}\propto u_{\text{imp}}/\gamma$ for small values of $u_{\text{imp}}$ and $n_{\text{imp}}$. Namely, the intrinsic anomalous Hall effect becomes more important in this case, and the extrinsic-intrinsic crossover becomes clearer.

\subsection{Extrinsic-intrinsic crossover}
\label{subsec:AHE:crossover}

\begin{figure}
  \begin{center}
    \includegraphics[width=7.2cm]{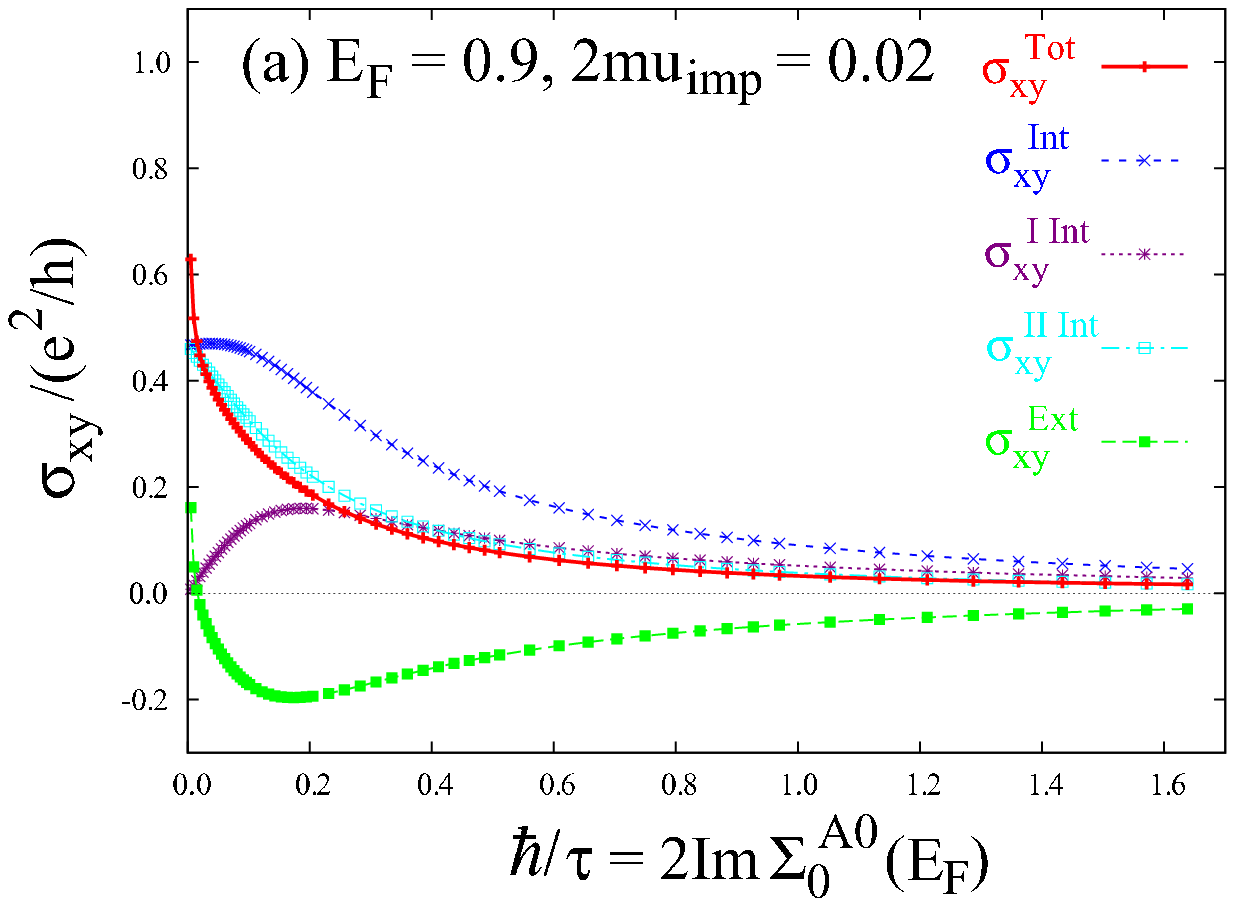}
    \includegraphics[width=7.2cm]{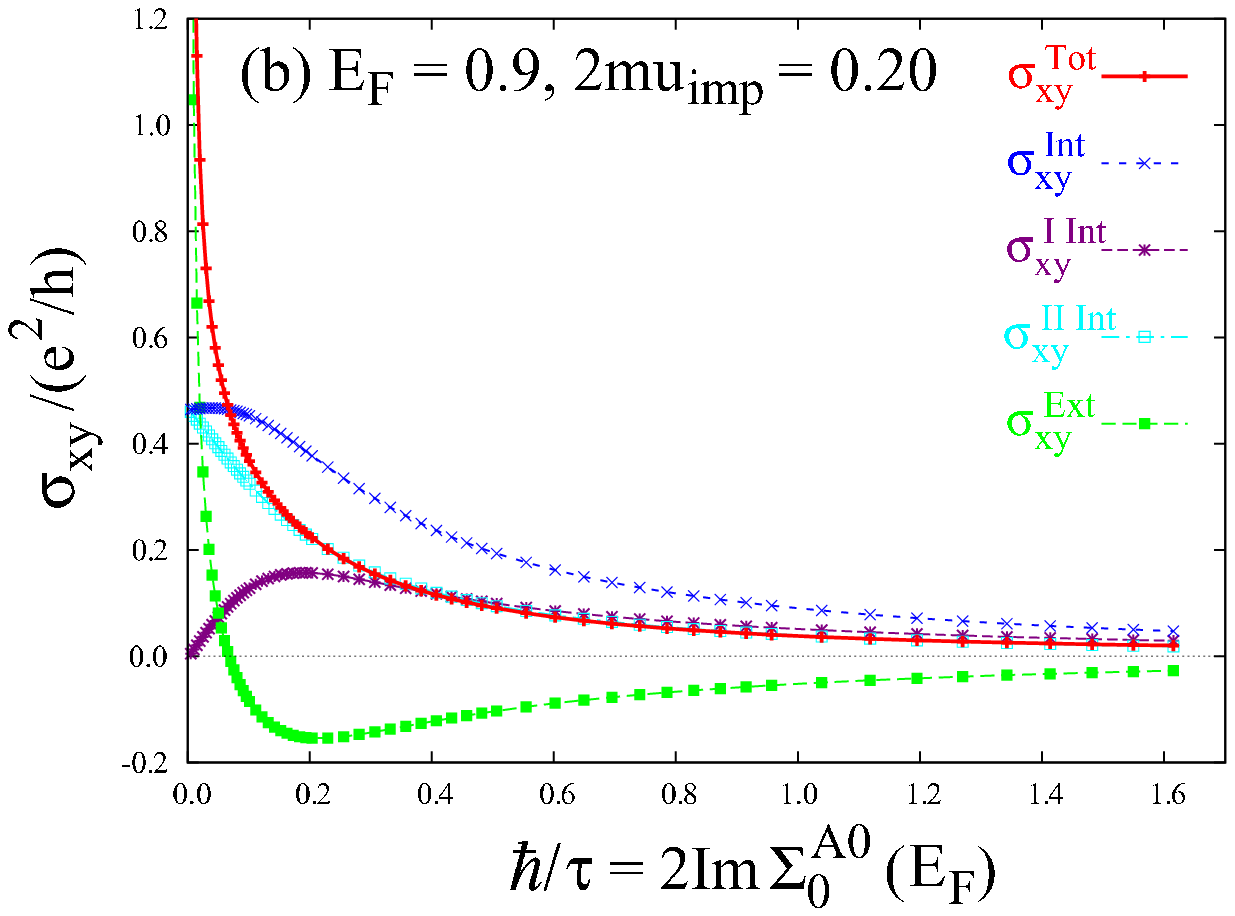}
    \includegraphics[width=7.2cm]{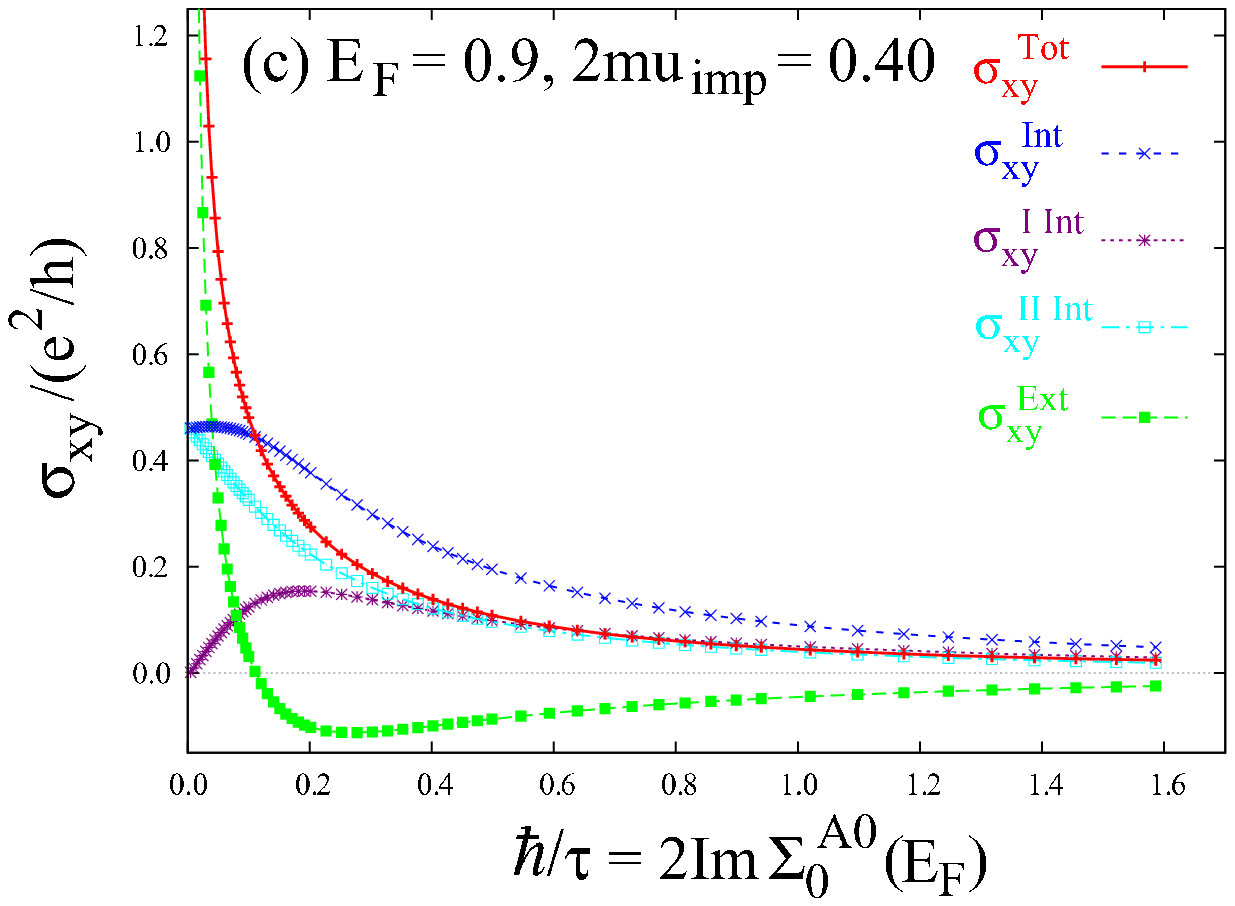}
    \includegraphics[width=7.2cm]{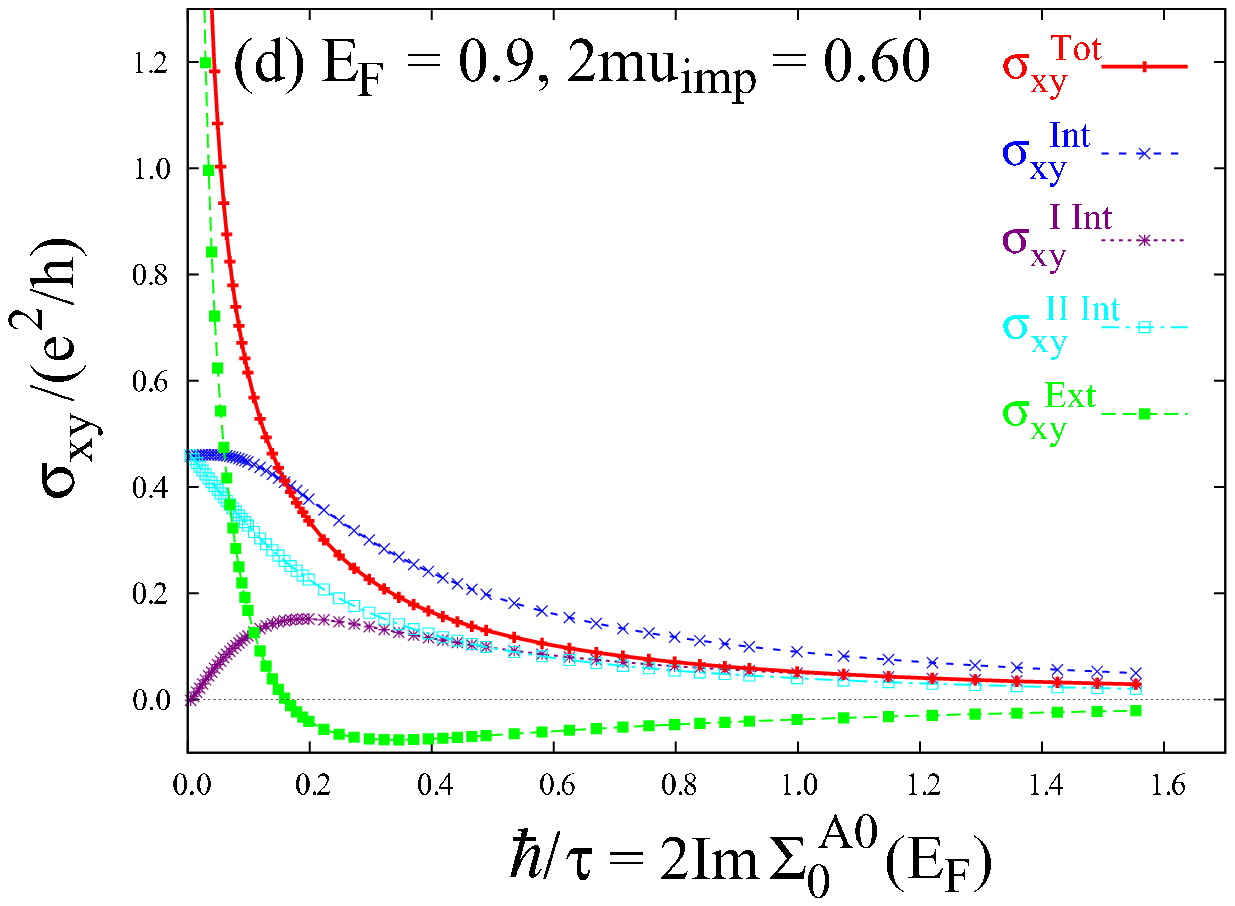}
  \end{center}
  \caption{\label{fig:theory1}(Color online) $\sigma_{xy}^\text{Tot}$, $\sigma_{xy}^\text{Int}$,$\sigma_{xy}^{I\ \text{Int}}$, $\sigma_{xy}^{II\ \text{Int}}$, and $\sigma_{xy}^\text{Ext}$ at $E_F=0.9$ as functions of the $\hbar/\tau$ for the same parameter values as Fig.~\ref{fig:Hall1} except the impurity potential strength (a) $2mu_{\text{imp}}=0.02$, (b) 0.2, (c) 0.4, and (d) 0.6.}
\end{figure}

\begin{figure}
  \begin{center}
    \includegraphics[width=7.2cm]{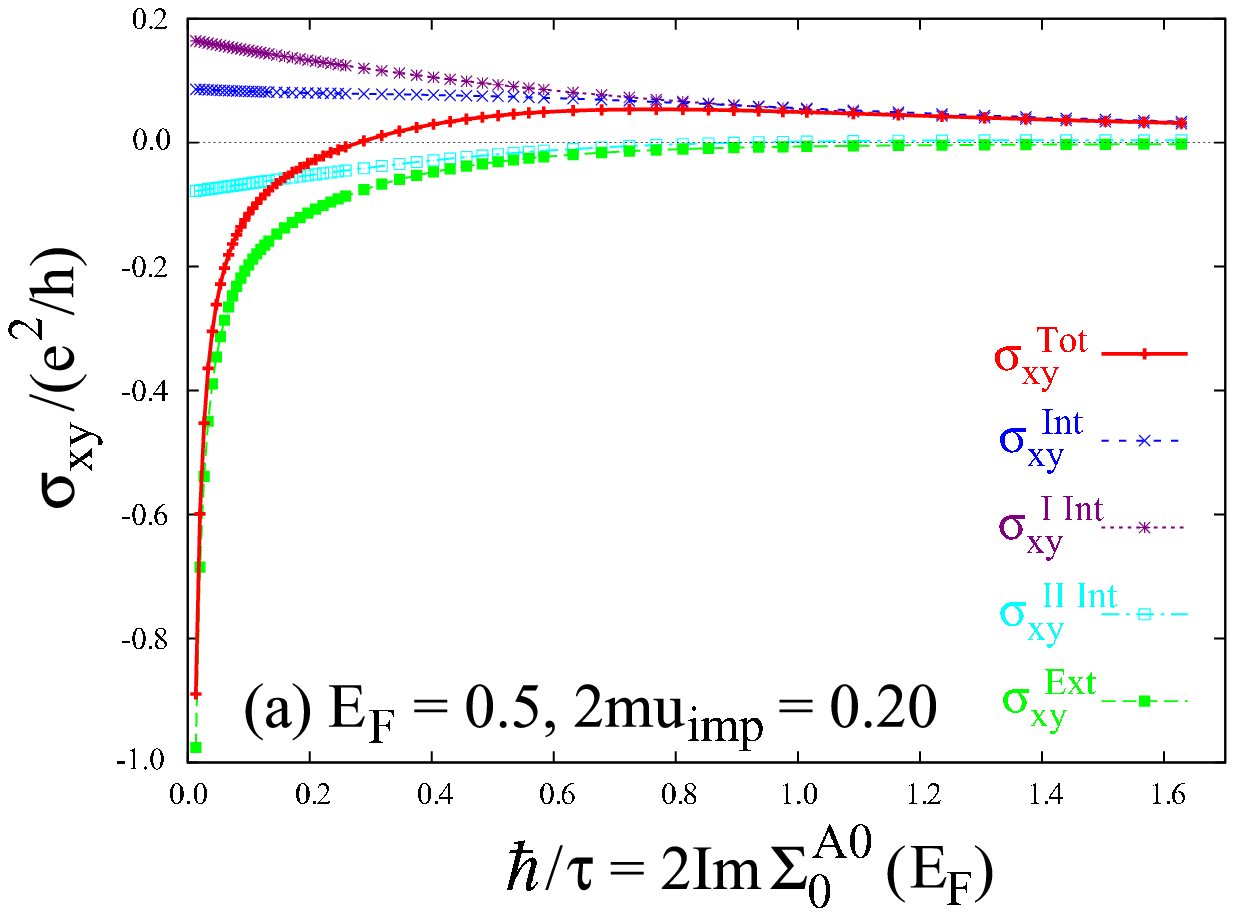}
    \includegraphics[width=7.2cm]{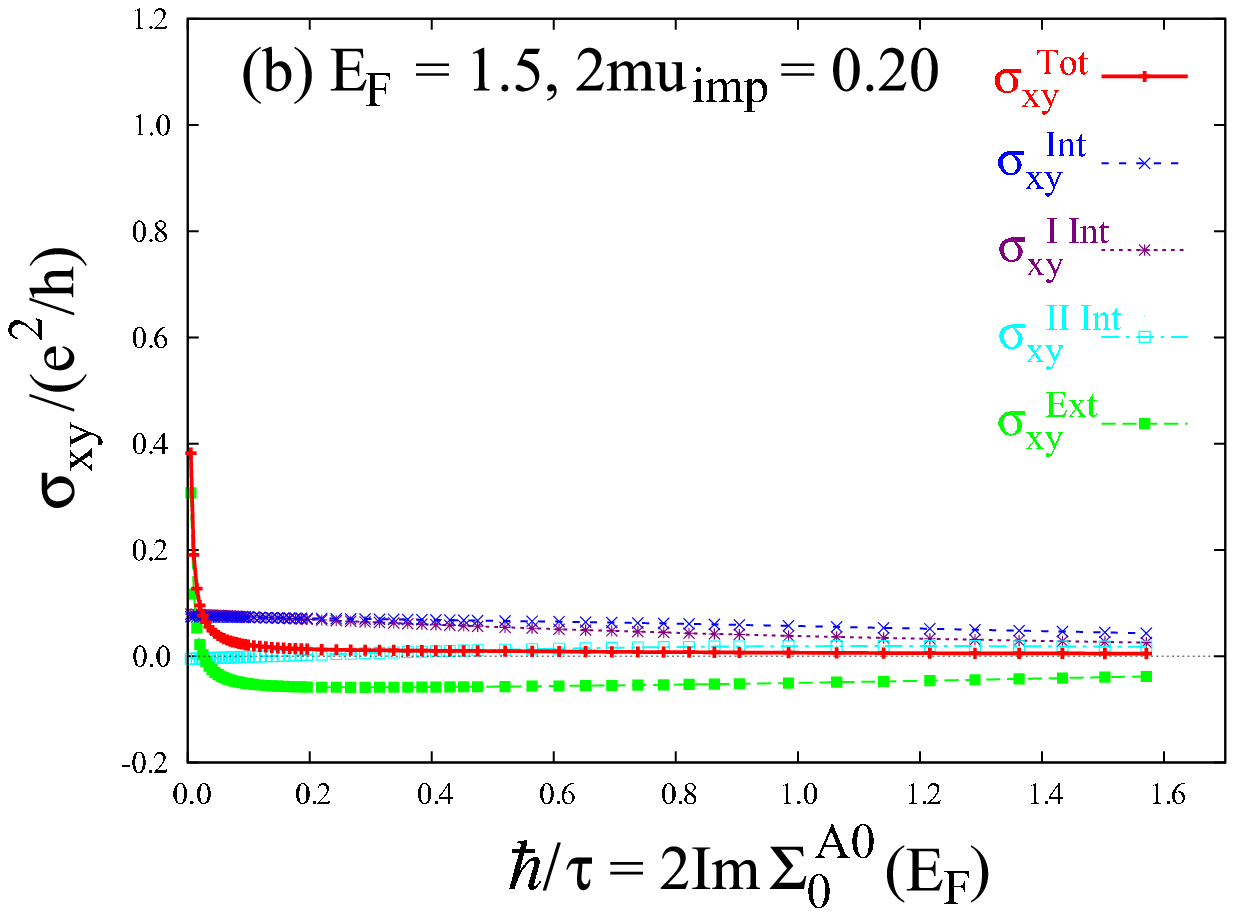}
  \end{center}
  \caption{\label{fig:theory2}(Color online) $\sigma_{xy}^\text{Tot}$, $\sigma_{xy}^\text{Int}$,$\sigma_{xy}^{I\ \text{Int}}$, $\sigma_{xy}^{II\ \text{Int}}$, and $\sigma_{xy}^\text{Ext}$ as a function of the $\hbar/\tau$ for the same parameter values as Fig.~\ref{fig:Hall1} with $E_F=0.5$, $0.9$ and $1.5$ for (a), (b), and (c), respectively.}
\end{figure}

Next, to gain a further insight and to see more clearly the extrinsic-intrinsic crossover in the anomalous Hall effect, we discuss $\sigma_{xy}^{\text{Int}}$, $\sigma_{xy}^{I\ \text{Int}}$, $\sigma_{xy}^{II\ \text{Int}}$, and $\sigma_{xy}^{\text{Ext}}$, in comparison with the total value $\sigma_{xy}^{\text{Tot}}$. Figures~\ref{fig:theory1} and \ref{fig:theory2} show the results for several choices of $E_F$ and $2mu_{\text{imp}}$ as a function of the averaged relaxation rate $\gamma=\hbar/\tau=2\text{Im}\Sigma^{A0}_0(E_F)$ at the Fermi level, instead of the Born scattering amplitude $\gamma_B$. Here, $\Sigma^{A0}_0(\varepsilon)$ is defined by Eq.~(\ref{eq:Sigma:4}) in Appendix~\ref{app:notation}. The other parameter values are taken to be the same as used in Fig.~\ref{fig:Hall1}. 

Let us start with the resonant case with $E_F$ being located within the resonant window. We show the results for $E_F=0.9$ (see the arrow b in Fig.~\ref{fig:dispersion}) in Fig.~\ref{fig:theory1}. It is clear that the extrinsic contribution $\sigma_{xy}^{\text{Ext}}$ and thus the total Hall conductivity $\sigma_{xy}^{\text{Tot}}$ include the component diverging in the clean limit $\tau\to\infty$, and the strength of the divergence is increased as the impurity potential strength $2mu_{\text{imp}}$ is varied from (a) 0.02, (b) 0.2, (c) 0.4, and (d) 0.6. This agrees with the skew-scattering scenario $\sigma_{xy}^{\text{skew}}\sim u_{\text{imp}}E_{\text{SO}}D\tau$ as given by Eq.~(\ref{eq:skew}). As noted in Sec.~\ref{subsec:AHE:E_F,tau}, the intrinsic contribution $\sigma_{xy}^{\text{Int}}$ as well as its components $\sigma_{xy}^{I\ \text{Int}}$ and $\sigma_{xy}^{II\ \text{Int}}$ are almost unchanged by this variation in $2mu_{\text{imp}}$. $\sigma_{xy}^{\text{Int}}$ has a nearly saturated value $\sim e^2/2h$ when $\hbar/\tau\lesssim\Delta_0=0.1$. 
Increasing the impurity scattering rate $\gamma=\hbar/\tau$ from the clean limit, $\sigma_{xy}^{I\ \text{Int}}$ evolves from 0, reaches the maximum, and then gradually decays, while $\sigma_{xy}^{II\ \text{Int}}$ monotonically decreases with increasing $\gamma$. Accordingly, the total intrinsic contribution is also only gradually decays as a function of $\gamma$. On the other hand, $\sigma_{xy}^{\text{skew}}$ rapidly decays in proportion to $\tau$. Then, as is clear from the panels (b)-(d) of Fig.~\ref{fig:theory1},  $|\sigma_{xy}^{\text{Int}}|$ and $|\sigma_{xy}^{\text{Ext}}|$ intersect at a value of the relaxation rate $\hbar/\tau$ proportional to $u_{\text{imp}}$. This is consistent with the semi-classical arguments: $\sigma_{xy}^{\text{Ext}}\sim\sigma_{xy}^{\text{skew}}\sim (e^2/h)u_{\text{imp}}E_{\text{SO}}D\tau/\hbar$ becomes comparable to $\sigma_{xy}^{\text{Int}}\sim e^2/2h$ at $\hbar/\tau\sim u_{\text{imp}}E_{\text{SO}}D$, and even smaller than $\sigma_{xy}^{\text{Int}}$ with futher increasing $\gamma=\hbar/\tau$. This defines the crossover bewtween the extrinsic anomalous Hall effect in the superclean system $\hbar/\tau \lesssim u_{\text{imp}}E_{\text{SO}}D$ and the intrinsic anomalous Hall effect in the moderately dirty system $\hbar/\tau \gtrsim u_{\text{imp}}E_{\text{SO}}D$. It should also be noticed that even within this intrinisc regime, another extrinsic contribution is present because of the vertex correction, i.e., $\hat{\Sigma}^<_{E_y,I}$, which partially cancels $\sigma_{xy}^{\text{Int}}$ and hence reduces $\sigma_{xy}^{\text{Tot}}$ from $\sigma_{xy}^{\text{Int}}$, as shown in Fig.~\ref{fig:theory1}.

For a small ratio of $E_{\text{SO}}/E_F\sim 10^{-3}-10^{-2}$ as in first-principles calculations~\cite{Fang03,Yao04} and $u_{\text{imp}}D\sim 1$, dominance of the intirnsic anomalous Hall effect is realized within the usual clean metal of $\hbar/\tau\gtrsim E_{\text{SO}}$ (several tens of meV). In reality, the total Hall conductivity is the sum of the contributions from all over the Brillouin zone. Since skew-scattering contributions from other momentum regions are always subject to a similar rapid decay, the above extrinsic-to-intrinsic crossover still occurs unless contributions from all the avoided-crossing regions of band dispersions are mutually canceled out by accident. 

Next, fixing the potential strength as $2mu_{\text{imp}}=0.2$, we take different values of the Fermi energy well below and above the resonant window. To be explicit, in the panels (a) and (b) of Fig.~\ref{fig:theory2}, we plot the results for $E_F=0.5$ and 1.5 marked with the arrows a and c in Fig.~\ref{fig:dispersion}, respectively. In both cases, the intrinsic contribution is much reduced from $e^2/2h$ by a factor of $2\Delta_0/E_{\text{res}}$. Therefore, the perturbation of $\sigma_{xy}$ in the spin-orbit interaction energy $E_{\text{SO}}$ is allowed in these cases. For (a) $E_F=0.5$, the dominance of the extrinsic skew-scattering contribution extends over a wide regime compared with the case of $E_F=0.9$ shown in Fig.~\ref{fig:theory1} (b). This agrees with arguments given in the Introduction along the Luttinger's theory (see Eq.~(\ref{eq:Luttinger2})). But with further increase in the relaxation rate $\gamma=\hbar/\tau$, the total Hall conductivity $\sigma_{xy}^{\text{Tot}}$ nearly merges into the Fermi-surface part $\sigma_{xy}^{I\ \text{Int}}$ of the intrinsic contribution. On the contrary to the resonant case shown in Fig.~\ref{fig:theory1}, not only $\sigma_{xy}^{II\ \text{Int}}$ but also $\sigma_{xy}^{I\ \text{Int}}$ is finite even in the clean limit. When the Fermi level is located above the resonant window, i.e., $E_F>\varepsilon_-(p=0)$, $\sigma_{xy}^{II\ \text{Int}}$ vanishes in the clean limit as shown in Fig.~\ref{fig:theory2} (b) for $E_F=1.5$, in agreement with the Kubo-formula calculation~\cite{Sinova}, and only gradually evolves into a finite value in the presence of a finite damping, which results from the self-consistency between the equilibrium Green's function $\hat{G}_0^{R,A}$ and self-energy $\hat{\Sigma}_0^{R,A}$ in the $T$-matrix approximation. $\sigma_{xy}^{I\ \text{Int}}$ is finite but it suffers from the reduction due to the vertex correction associated with $\hat{\Sigma}_{E_y,I}^<$. Then, the total Hall conductivity $\sigma_{xy}^{\text{Tot}}$ nearly vanishes except the skew-scattering contribution.

\subsection{Scaling relations between $\sigma_{xx}$ and $\sigma_{xy}$}\label{subsec:scaling}

\begin{figure}
  \begin{center}
    \includegraphics[width=7.2cm]{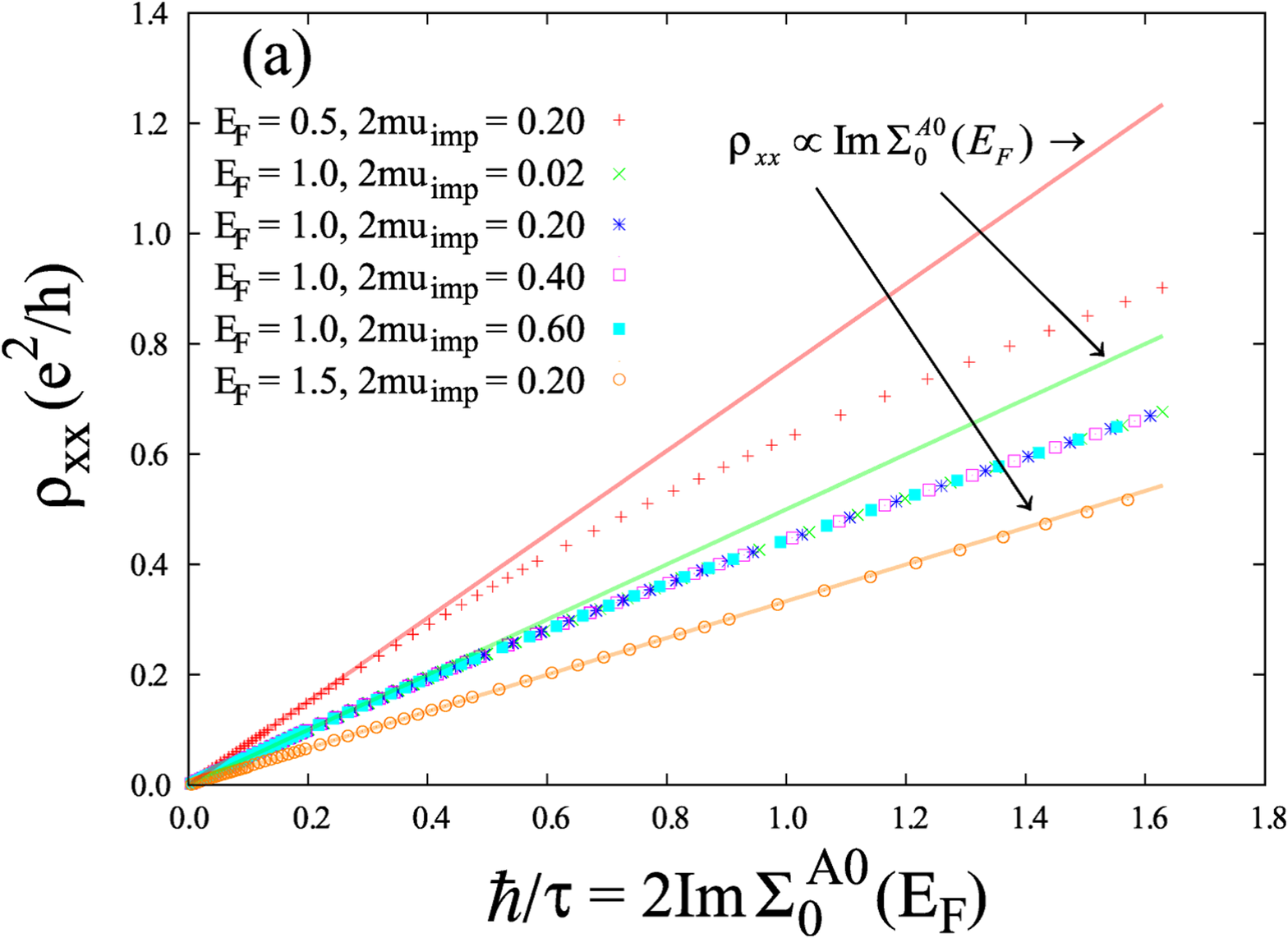}
    \includegraphics[width=7.2cm]{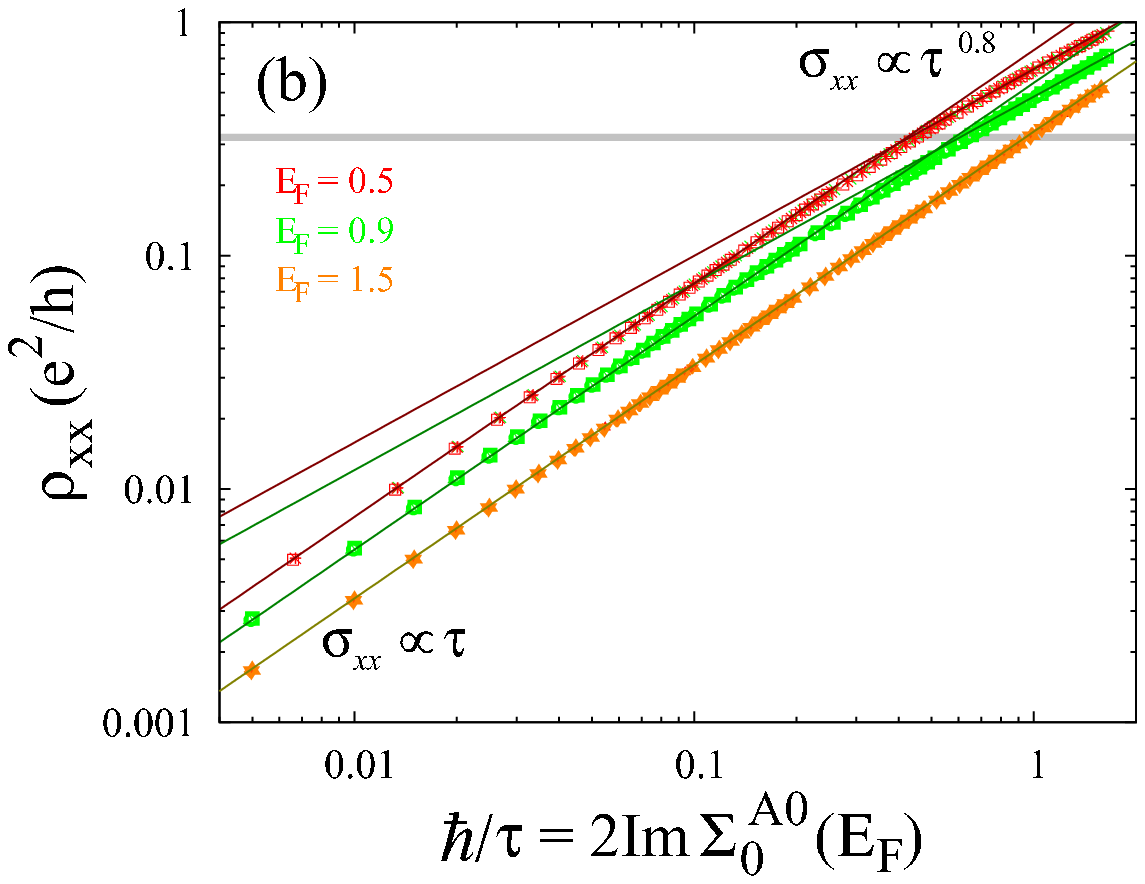}
  \end{center}
\caption{\label{fig:resistivity} (Color online) Resistivity $\rho_{xx}\sim1/\sigma_{xx}$ for the same set of parameters as Fig.~\ref{fig:Hall1} except a variation of the impurity potential strength $2mu_{\text{imp}}=0.02$, $0.2$, $0.4$ and $0.6$. (a) and (b) are the linear and logarithmic plots. In (b) all the data points for $2mu_{\text{imp}}=0.02$, 0.2, 0.4, and 0.6 are plotted with the same symbol for each value of $E_F$.}
\end{figure}

In the rest of this section, we discuss scaling relations between $\sigma_{xy}$ and $\sigma_{xx}$, which is a source of the controversy on the interpretation of the experimental results.

First, we show the results of the resistivity $\rho_{xx}\sim1/\sigma_{xx}$ calculated by using the same parameter values $2mv^2=3.59$ and $\Delta_0=0.1$ in Fig.~\ref{fig:resistivity}. The data with different values of $2mu_{\text{imp}}$ fall into a single curve for each $E_F$. In the clean limit, $\rho_{xx}$ is proportional to $\gamma=\hbar/\tau$ or equivalently $\sigma_{xx}$ is proportional to $\tau$ with different coefficients depending on the Fermi energy $E_F$. However, it is clear from both the linear and logarithmic plots in (a) and (b), respectively, that for $E_F\lesssim E_{\text{res}}$, the resistivity exhibits a different scaling relation $\rho_{xx}\propto(\hbar/\tau)^{\varphi_{xx}}$ with $\varphi_{xx}\sim0.8\pm0.05$, as in the portion of the $E_F=0.5$ and 0.9 curves. This crossover occurs as the resistivity is varied across $h/\pi e^2$.

\begin{figure}
  \begin{center}
    \includegraphics[width=7.2cm]{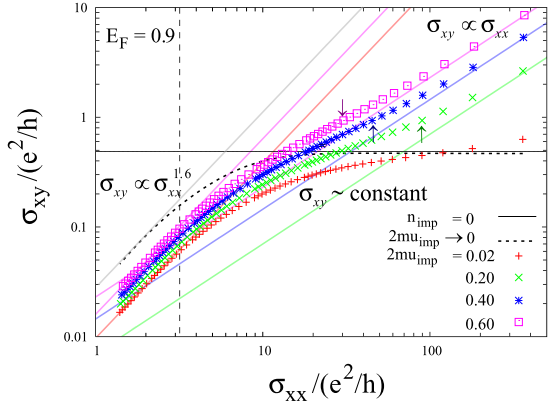}
  \end{center}
  \caption{ (Color online) Scaling plot of $\sigma_{xy}$ versus $\sigma_{xx}$ for the same sets of parameter values as in Fig.~\ref{fig:theory2}(b) except $2mu_{\text{imp}}$. The arrows show the extrnsic-intrinsic crossover scale of $\sigma_{xx}$ for $2mu_{\text{imp}}=0.20$, 0.40, and 0.60. The horizontal solid line and the dotted curve represent the values given by the TKNN formula of Eq.~(\ref{eq:TKNN}) and the intrinsic contribution in the limit of $2mu_{\text{imp}}\to0$. The vertical dashed line gives the second crossover scale of $\sigma_{xx}(\sim \pi e^2/h)$ to the 1.6 power-law regime in the dirty regime.}
\label{fig:scaling}
\end{figure}

 Figure~\ref{fig:scaling} shows the logarithmic plot of $\sigma_{xy}$ against $\sigma_{xx}$ for the same set of parameters as Fig.~\ref{fig:theory1}. In the clean limit, the curves nicely follow $\sigma_{xy}\propto\sigma_{xx}$ and the ratio $\sigma_{xy}/\sigma_{xx}$ is proportional to $u_{\text{imp}}$ for a fixed $\tau$ or $\sigma_{xx}$. Note, however, that for $2mu_{\text{imp}}=0.02$, the intrinsic anomalous Hall effect is still robust within this range of $\sigma_{xx}$, as argued in Sec.~\ref{subsec:AHE:E_F,tau}. As $\sigma_{xx}\sim2(e^2/h)E_F\tau/\hbar$ decreases with decrease in $\tau$, the relation between $\sigma_{xx}$ and $\sigma_{xy}$ exhibits a upward deviation from the linear one $\sigma_{xy}\propto\sigma_{xx}$, signalling the crossover to the intrinsic regime with an almost constant $\sigma_{xy}$. 
As we have already mentioned, this extrinsic-intrinsic crossover occurs around $\sigma_{xx} \sim (e^2/ha) (E_F/u_{\text{imp}}E_{\text{SO}}D)$.
In terms of the resistivity tensor, the crossover occurs when $\rho_{xx}$ is of the order of $\mu\Omega$~cm; $\rho_{xy}\propto\rho_{xx}$ in more conducting region, while $\rho_{xy}\propto\rho_{xx}^2$ in less conducting region.

A smaller impurity potential strength $u_{\text{imp}}$ enlarges the region of the constant behavior of $\sigma_{xy}$. (Note that we change $n_{\rm imp}$ to control $\hbar/\tau$.) It is remarkable that in the case of $2mu_{\text{imp}}=0.02$, the intrinsic behavior of an almost constant $\sigma_{xy}$ is clearly observed in the moderately diry case. This intrinsic regime with an almost constant $\sigma_{xy}$ continues even with the variation of $\sigma_{xx}$ over three orders of magnitude. The magnitude of $\sigma_{xy}$ in the intrinsic regime is also consistent with experimentally observed values $\sigma_{xy} \cong 10^2-10^3 \ \Omega^{-1}{\rm cm}^{-1}$ in this $\sigma_{xy}$-constant region, as is summarized in Sec.~\ref{sec:exp}.

Further decrease in $\tau$ changes the scaling behaviour to $\sigma_{xy}\propto(\sigma_{xx})^{\varphi}$ with $\varphi\sim1.6$. This nontrivial exponent appears in the dirty regime $\sigma_{xx}\lesssim\pi e^2/h$, where the longitudinal transport also exhibits the nontrivial scaling $\sigma_{xx}\propto\gamma^{0.8}$ and hence the Hall conductivity scales as $\sigma_{xy}\propto \tau^{\varphi_{xy}}$ with $\varphi_{xy}=\varphi\varphi_{xx}\sim1.3$. This exponent appears irrespective of the position of the Fermi level. Actually, the exponent $\varphi_{\sigma}$ has recently been experimentally verified in various class of ferromagnetic materials, as is summarized in Sec.~\ref{sec:exp}. Note also that this exponent approximates to the value expected for the insulating phase of the quantum Hall system under a strong applied magnetic field~\cite{PryadkoAuerbach04}. These nonperturbative exponents are obtained as a result of the self-consistency between the Green's function and the self-energy and the inclusion of the vertex correction. The intuitive understanding requires a further study. It should also be noticed that the present theory includes neither the weak localization corrections steming from the Cooperons and diffusons nor effects of the Anderson localization, which are crucial to explain recent nontrivial experimental observations on disordered thin films~\cite{Hebard}.

\section{Anomalous Nernst effect}\label{sec:ANE}

Before proceeding to discussions and conclusions, we give numerical results for the anomalous Nernst effect. The $T$-linear coefficient of the thermoelectric Hall conductivity $\alpha_{xy}$ is calculated by evaluating the expressions Eqs.~(\ref{eq:alpha}), (\ref{eq:alpha^I}), and (\ref{eq:alpha^II}) obtained through the Mott rule. This includes information on the topological structure of the wavefunction at the Fermi level, which can not be directly gained from an observation of $\sigma_{xy}$.  

\begin{figure}
  \begin{center}
    \includegraphics[width=8.0cm]{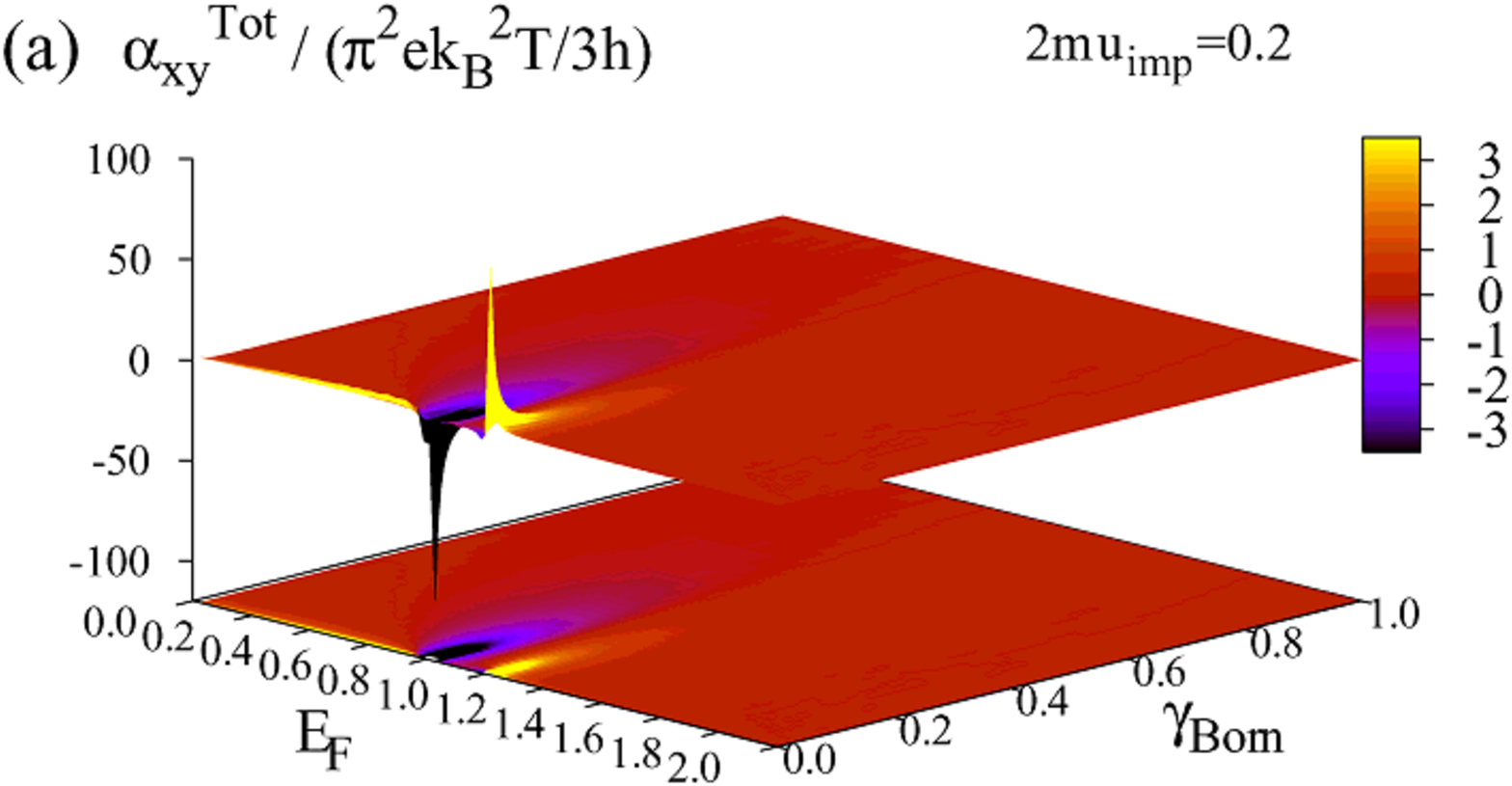}
    \includegraphics[width=8.0cm]{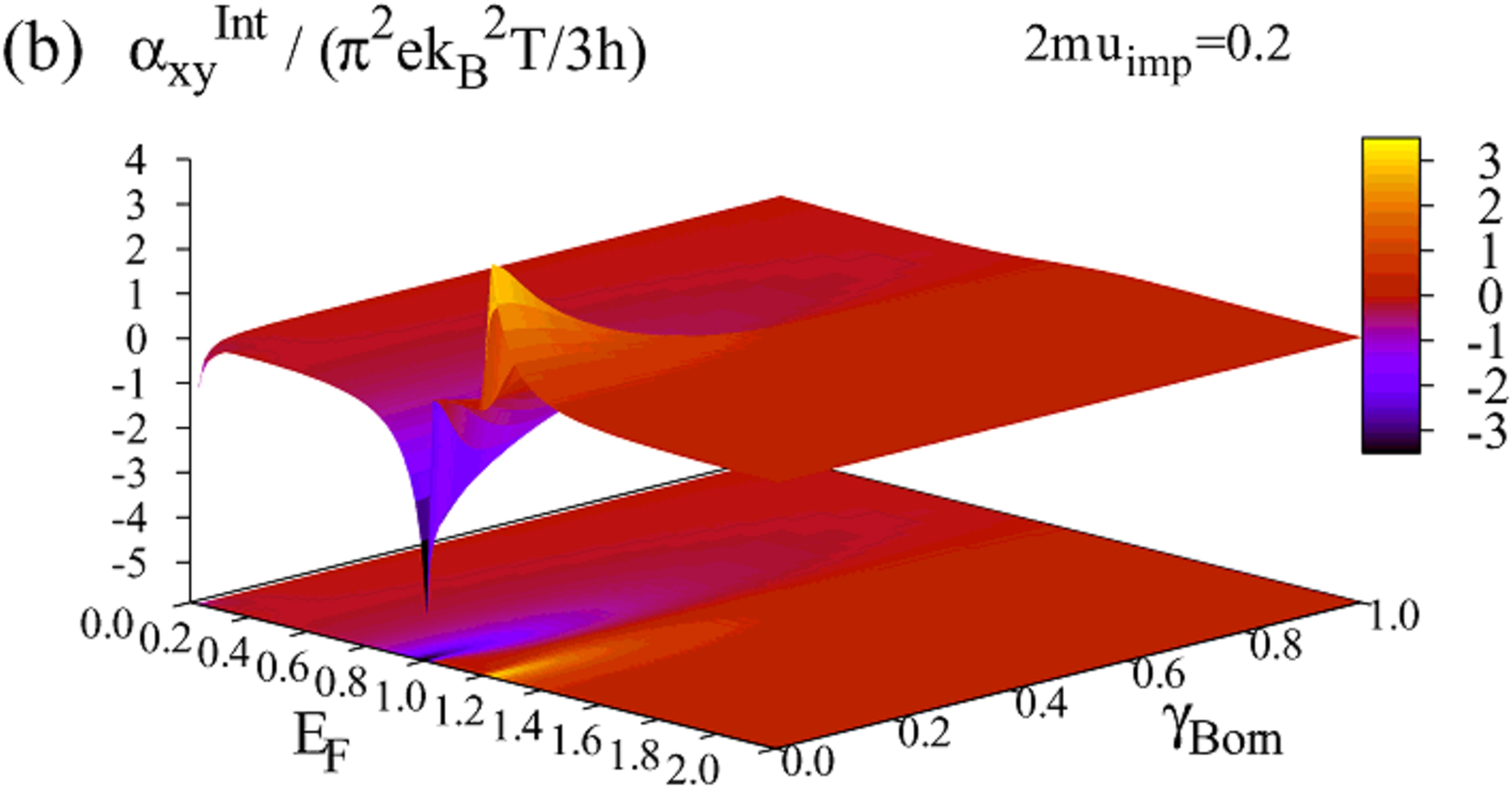}
  \end{center}
  \caption{(Color online) (a) The $T$-linear coefficients of the total anomalous thermoelectric Hall conductivity $\alpha_{xy}^{\text{Tot}}/T$ and (b) the intrisic contribution $\alpha_{xy}^{\text{Int}}/T$ as functions of $E_F$ and $\hbar/\tau$ in an energy unit of $E_{\rm res}=1.0$. The parameters are choisen as $v=3.59$, $\Delta_0=0.1$, and $2mu_{\text{imp}} = 0.2$, as in Fig.~\ref{fig:Hall1}. Note the difference of the scales for $\alpha_{xy}$ in (a) and (b).}
\label{fig:Nernst1}
\end{figure}

First, $\alpha_{xy}^{\text{Tot}}/T$ and $\alpha_{xy}^{\text{Int}}/T$ for the total and the intrinsic anomalous Nernst effects are shown as a function of the Fermi energy $E_F$ and the Born scattering amplitude $\gamma_{\text{Born}}$ in the panels (a) and (b) of Fig.~\ref{fig:Nernst1}, respectively. Here, the same parameter values as in Fig.~\ref{fig:Hall1} have been used. In the clean limit, the skew-scattering conrtibution $\alpha_{xy}^{\text{skew}}$ yields a dominant and diverging contribution for $E_F\lesssim \varepsilon_+(p=0)=E_{\text{res}}-2\Delta_0=0.8$ (see Fig.~\ref{fig:dispersion}), reflecting the $E_F$ dependence of $\sigma_{xy}^{\text{skew}}$ as found in Fig.~\ref{fig:Hall1} (a) (or (c)). Besides, there appear two prominent structures at $E_F=\varepsilon_\pm(p=0)$, where the sign change of $\sigma_{xy}^{\text{skew}}$ occurs and hence the $\alpha_{xy}$ is also strongly enhanced at these points with opposite signs. On the other hand, $\alpha_{xy}$ is appreciably suppressed in the resonant window $E_F\in[E_{\text{res}}-2\Delta_0,E_{\text{res}}]$. With increasing $\gamma_{\text{Born}}$, $\alpha_{xy}^{skew}$ rapidly decays, as in the case of the anomalous Hall effect. Figure~\ref{fig:Nernst1} (b) shows the intrinsic contribution $\alpha_{xy}^{\text{Int}}$ calculated by imposing the condition of $\hat{\Sigma}_{E_y}^{R,A,<}=0$, as in the case of $\sigma_{xy}^{\text{Int}}$. In the pure case $n_{\text{imp}}=0$, $\alpha_{xy}^{\text{Int}}$ is proportional to the density of the Berry-phase curvature at the Fermi level. This shows that the Berry-phase curvature is strongly enhanced around $E_F=\varepsilon_\pm(p=0)$, and that it only gradually decays as a function of the relaxation $\gamma_{\text{Born}}$. Then, in moderately dirty systems, there appears a crossover from the extrinsic skew-scattering regime to the intrinsic regime. However, in the case of the anomalous Nernst effect, an interference between positive and negative contributions appears and a sign change of $\alpha_{xy}$ often occurs even as a function of $\gamma$. Therefore, detailed scaling analysis as performed for $\sigma_{xy}$ is difficult.

Note that the divergence of $\alpha_{xy}/T$ emerges only in the clean limit, in particular, around $E_F=\varepsilon_\pm(p=0)$. In the clean limit, $\sigma_{xy}^{\text{Int}}$ exhibits a kink at $E_F=\varepsilon_\pm(p=0)$. In the presence of finite relaxation, the quasiparticle spctra are broadened and this smears out any singularity. However, if the broadening effect is ignored, then $\sigma_{xy}^{\text{Tot}}$ shows a discontinuity at $E_F=\varepsilon_-(p=0)$, namely when the chemical potential crosses the bottom of the minority band, no matter how large is the relaxation associated with the impurity scattering~\cite{Sinova}. According to the Mott rule given by Eq.~(\ref{eq:Mott}), this results in a $\delta$-functional divergence $\alpha_{xy}/T\to\infty$ at this Fermi energy. Therefore, one must take into account the finite damping of quasipaticles in performing the energy-momentum integration to calculate the conductivity tensor.

\section{Relation with other theories}\label{sec:others}

In this section, we clarify the relations of the present theory to recent and early theories.

Recently, the anomalous Hall conductivity in this model or similar models have also been intensively discussed using the Kubo and/or the Streda formulae~\cite{Dugaev05,Sinova,Inoue} and the effective semi-classical Boltzmann transport theory~\cite{Sinova}. Some of them ignored the vertex corrections totally~\cite{Kontani} or partially~\cite{Inoue}. Others neglected the self-energy corrections or just took a constant relaxation-time approximation~\cite{Sinova,Inoue,Kontani}, and further assumed that the relaxation rate is independent of the band index~\cite{Inoue}, which is justified in the limit of $E_F\gg \varepsilon_-(p=0)$. Then, they gave partly different behaviors from our results shown in the previous Letter~\cite{Onoda06_prl} and the present paper. We stress that the vertex corrections are crucial for the transport properties and gives a significantly important contribution particularly in the clean limit, namely, the skew-scattering contribution. In the dilute impurity concentration, the scattering potential has a particle-hole asymmetry and gives rise to the asymmetric scattering together with the spin-orbit interaction for the Bloch electrons, even without the potential having a large spin-orbit interaction studied in the original work~\cite{Smit55,Smit58} or $f$-electrons~\cite{Coleman85}. Note that the lack of the self-consistency between the equilibrium Green's function and self-energy in the litrature~\cite{Inoue,Sinova} is another reason for the discrepancy. 

Now the main disagreement and the controversy on the anomalous Hall effect in this model described by Eq.~(\ref{eq:H}) is on the absence or the existence of the skew-scattering contribution and the total anomalous Hall conductivity in the case where the both bands are partially filled, as addressed by Sinova and coworkers~\cite{Sinova}. There are two main sources for this discrepancy. In Ref.~\onlinecite{Sinova}, (i) the self-energy correction was taken as only two constant relaxation rates in the Born approximation, whereas we have solved the self-consistent $T$-matrix approximation which becomes exact in the dilute impurity limit. (ii) When momentum integrations of the Green's functions were performed to calculate $\sigma_{xy}$ by means of the Kubo formula, they employed the semi-classical approximation where the quasipaticle spectrum had a $\delta$-functional form and hence neglected the self-energy in the denominator. Such expansion of $\sigma_{xy}$ at the singular point $\tau\to\infty$ is not straightforward to handle in the present model. In Ref.~\onlinecite{Onoda06_prl} and the present paper, we have taken the opposite strategy free from the singularity: we start from the case with a finite lifetime broadening due to the impurity scattering, and gradually decrease the impurity scattering strength. This discrepancy on the anomalous Hall effect becomes striking when we consider the anomalous Nernst effect, as we mentioned in Sec.~\ref{sec:ANE}. Namely, applying the Mott rule to the singular results obtained for $\sigma_{xy}$ in Ref.~\onlinecite{Sinova}, the $T$-linear coefficient to the thermoelectric Hall conductivity tensor $\alpha_{xy}$ diverges when the Fermi level crosses the bottom of the upper band, i.e., at $E_F=\varepsilon_-(p=0)$, even with the finite relaxation rate. This is not plausible. From the viewpoints of the semi-classical Boltzmann transport theory and the Kubo-formula calculation in the Matsubara technique, it is required to modify the calculations of these corrections beyond the semi-classical approximation. The above arguments indicate that both the vertex corrections and the self-energy corrections are highly important for the transport properties and should be properly taken into account. 

The present theory explains the anomalous Hall effect in the whole regime except in the localized regime. From the present results, now the source of the confusion over decades is clear. The amplitude of the skew-scattering contribution, though it is rather sensitive to details of the impurity potential and band structure, can be larger than $e^2/h$ in the superclean case $\hbar/\tau\ll E_{\text{SO}}$, if we assume the impurity potential strength of the order of the bandwidth or the Fermi energy. In this case, there is no chance for the band calculation to reproduce the observed value of $\sigma_{xy}$ and it is difficult to explain the anomalous Hall effect quantitatively from the theoretical viewpoint. On the other hand, this skew-scattering contribution decays for $E_{\text{SO}} \lesssim \hbar/\tau$ rapidly. The side-jump contribution is also small and of the order of $(e^2/h)(E_{\text{SO}}/E_F)$~\cite{Nozieres73}. Thefore, the intrinsic contribution, which is of the order of $e^2/h$ under the resonant condition, is dominant over a wide range of the scattering strength $E_{\text{SO}}\lesssim\hbar/\tau\lesssim E_F$ (clean or moderately dirty case). Although Luttinger reconsidered the Karplus-Luttinger theory~\cite{KarplusLuttinger54} and gave an expansion of $\sigma_{xy}$ in $u_{\text{imp}}$, including the skew-scattering contribution as well~\cite{Luttinger58}, it fails to reveal the above crossover in the space of $E_F$, $E_{\text{SO}}$ and $\hbar/\tau$.

Our theory also confirms the condition for the first-principles band calculation of the intrinsic anomalous Hall conductivity to work reasonably in comparison with experiments. It can elucidate the experimentally oberved value of $\sigma_{xy}$ except a correction arising from a reduction due to the vertex correction, when the resonantly enhanced intrinsic anomalous Hall effect dominantly determines the $\sigma_{xy}$. This actually occurs in the moderately dirty case where $\sigma_{xy}$ only weakly depends on the scattering rate, as shown in Sec.~\ref{sec:AHE}. 

\section{Comparison with experiments}
\label{sec:exp}

\begin{figure*}[htb]
  \begin{center}
    \includegraphics[width=14.cm]{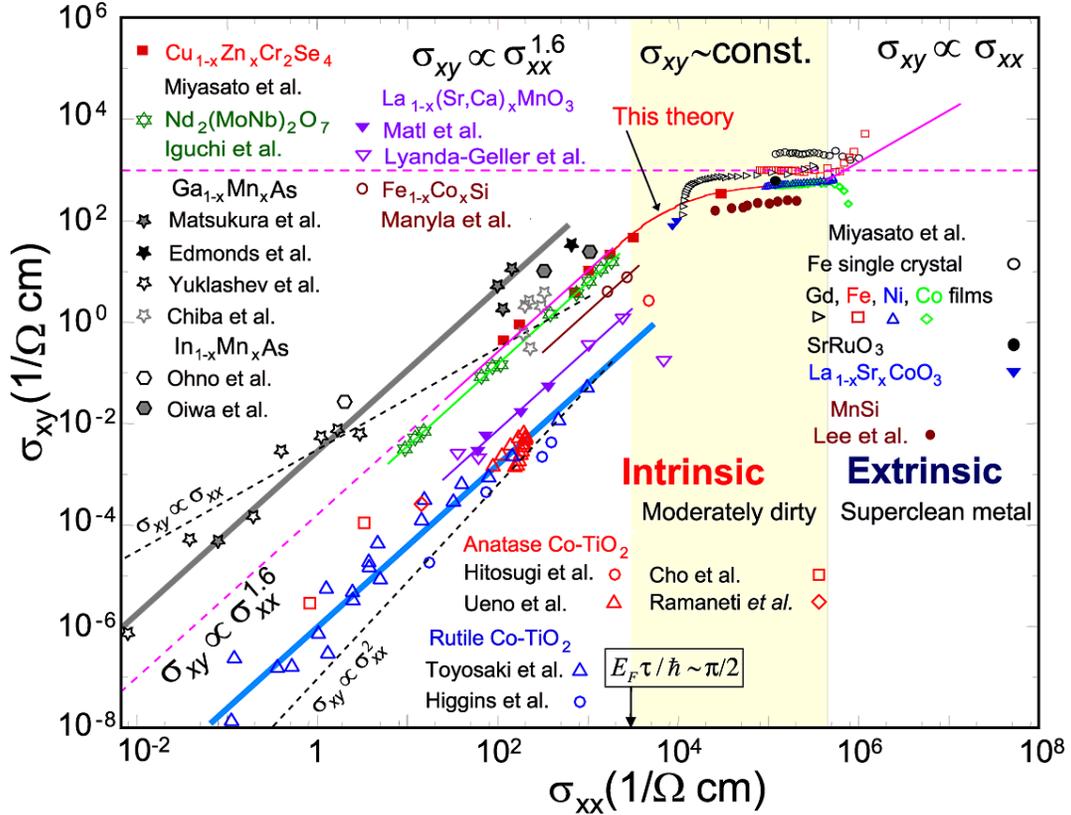}
  \end{center}
  \caption{(Color online) Summary of experimental results on various ferromangets, including transition-metals, perovskite oxides, spinels, and magnetic semiconductors. The theoretical curve corresponding to Fig.~\ref{fig:scaling} ($2mu_{\text{imp}}=0.02$) is also shown. The data are taken from Miyasato \textit{et al}.~\cite{Miyasato07} for Gd, Fe, Ni, and Co films, Fe single crystals, SrRuO$_3$, La$_{1-x}$Sr$_x$CoO$_3$ ($x=0.20$, 0.25) and Cu$_{1-x}$Zn$_x$Cr$_2$Se$_4$ ($x=0.0$, 0.2, 0.4, 0.5, 0.6, 0.8, 0.9), from Lee \textit{et al}.~\cite{Lee_science04} and Lyanda-Geller \textit{et al}.~\cite{Lyanda-Geller01} for La$_{1-x}$(SrCa)$_x$MnO$_3$, from Iguchi \textit{et al}.~\cite{Iguchi07} for Nd$_2$(MoNb)$_2$O$_7$, from Manyala \textit{et al}.~\cite{Manyala_nma04} for Fe$_{1-x}$Co$_x$Si, from Lee \textit{et al}.~\cite{Ong07} for MnSi, from Matsukura \textit{et al}.~\cite{Matsukura98}, Edmonds \textit{et al}.~\cite{Edmonds03}, Yuldashev \textit{et al}.~\cite{Yuldashev04}, and Chiba \textit{et al}.~\cite{Chiba07} for Ga$_{1-x}$Mn$_x$As, from Ohno \textit{et al}.~\cite{Ohno92} and Oiwa \textit{et al}.~\cite{Oiwa99} for In$_{1-x}$Mn$_x$As, from Ueno \textit{et al}.~\cite{Ueno06}, Cho \textit{et al}.~\cite{Cho06}, and Ramaneti \textit{et al}.~\cite{Ramaneti} for anatase-Co-TiO$_2$, and from Toyosaki \textit{et al}.~\cite{Toyosaki04} and Higgins \textit{et al}.~\cite{Higgins04} for rutile-Co-TiO$_2$. }
  \label{fig:exp}
\end{figure*}
We now turn to the comparison with experimental results. The anomalous Hall effect has been investigated as a fundamental property in many ferromagnetic materials with careful analyses to separate the anomalous component from the ordinary one. The results on $\sigma_{xx}$ and $\sigma_{xy}$ at low temperatures are summarized in Fig.~\ref{fig:exp} for Fe, Ni, Co, and Gd films~\cite{Miyasato07}, Fe single crystals~\cite{Miyasato07}, SrRuO$_3$~\cite{Miyasato07}, La$_{1-x}$Sr$_x$CoO$_3$~\cite{Miyasato07}, Cu$_{1-x}$Zn$_x$Cr$_2$Se$_4$~\cite{Miyasato07}, La$_{1-x}$(SrCa)$_x$MnO$_3$~\cite{Lyanda-Geller01,Lee_science04}, Nd$_2$(MoNb)$_2$O$_7$~\cite{Iguchi07}, Fe$_{1-x}$Co$_x$Si~\cite{Manyala_nma04}, MnSi~\cite{Ong07}, Ga$_{1-x}$Mn$_x$As~\cite{Matsukura98,Edmonds03,Yuldashev04,Chiba07}, In$_{1-x}$Mn$_x$As~\cite{Ohno92,Oiwa99}, anatase-Co-TiO$_2$~\cite{Ueno06,Cho06,Ramaneti}, and rutile-Co-TiO$_2$~\cite{Toyosaki04,Higgins04}. It is significantly important that all the experimental data are categorized into three regimes. 

In the poorly conducting regime, there exists a universal scaling relation of $\sigma_{xy}\propto\sigma_{xx}^{1.6}$, which agrees fairly well with the present theory. This is the case for Cu$_{1-x}$Zn$_x$Cr$_2$Se$_4$~\cite{Miyasato07}, La$_{1-x}$Sr$_x$CoO$_3$~\cite{Miyasato07}, a disordered pyrochlore ferromagnet Nd$_2$(Mo$_{1-x}$Nb$_x$)$_2$O$_7$~\cite{Iguchi07}, Co-doped TiO$_2$~\cite{Ueno06}, Mn-doped GaAs~\cite{Gossard,Fukumura07}. The difference in the amplitudes can be understood as a difference in the number of momentum regions with avoided-crossing and/or a difference in the relative position of the Fermi level. A naive interpretation in terms of $\sigma_{xy}\propto\sigma_{xx}^2$, which can be obtained by calculating $\sigma_{xy}$ from the perturbative expansion in the quasiparticle damping rate, neglecting the vertex corrections, and assuming that $\sigma_{xx}$ is proportional to the damping rate, gives a clearly worse fitting to the experimental data than $\sigma_{xy}\propto\sigma_{xx}^{1.6}$. Though another scaling behavior of $\sigma_{xy}\propto\sigma_{xx}$ might also be appropriate for some experimental data on Ga$_{1-x}$Mn$_x$As~\cite{Ohno}, $\sigma_{xy}\propto\sigma_{xx}^{1.6}$ can explain its global dependence. 

In the moderately dirty regime with $\sigma_{xx}\sim3\times10^3$-$5\times10^5$ $\Omega^{-1}$ cm$^{-1}$, $\sigma_{xy}$ has only a gradual dependence on $\sigma_{xx}$ and appears to approach constant values of the order of $10^2$-$10^3$ $\Omega^{-1}$ cm$^{-1}$ with increasing $\sigma_{xx}$, as observed in Fe- and Ni-based dilute alloys~\cite{Hurd}, Cu$_{1-x}$Zn$_x$Cr$_2$Se$_4$~\cite{Miyasato07}, SrRuO$_3$~\cite{Miyasato07}, metallic foils for Fe, Ni, Co, and Gd~\cite{Miyasato07}, and MnSi~\cite{Ong07}. The large amplitude and the robustness against the scattering events are consistent with the intrinsic scenario. The side-jump mechanism also yields an almost constant behavior of $\sigma_{xy}$. However, it suffers from a small factor of $E_{\text{SO}}/E_F\sim10^{-2}-10^{-1}$ coompared with the intrinsic Berry-phase contribution under the resonant condition, which can not account for the large amplitude $\sigma_{xy}\sim10^3$ $\Omega^{-1}$ cm$^{-1}$. Therefore, the experimental results in this regime can be mostly assigned to the intrinsic Berry-phase contribution. Actually, the theoretical curve based on the present simplest analysis with $E_F=0.9$, $2mv^2=3.59$, and $2mv_{\text{imp}}=0.02$, which is shown by the red curve, can even explain the gradual dependence of $\sigma_{xy}$ on $\sigma_{xx}$ for Cu$_{1-x}$Zn$_x$Cr$_2$Se$_4$ from the dirty to the moderately dirty regime. 

In the superclean regime with $\sigma_{xx}\gtrsim5\times10^6$ $\Omega^{-1}$ cm$^{-1}$, the curve for $\sigma_{xy}$ versus $\sigma_{xx}$ tends to deviate from the constant behavior, and there appears a rapid increase or decrease by another contribution, which is positive or negative to the intrinsic contribution, respectively, depending on meterials. The detailed data seem to depend on the properties of dilute impurities embedded in the materials. In fact, the ordinary Hall effect is pronounced by the Landau-level formation at low magnetic fields because of the high mobility. It gives a nonlinear dependence of the Hall resistivity on the applied magnetic field. Furthermore, the remnant magnetization tends to decrease with increasing the purity in many highly conducting materials as Fe and Ni. Therefore, the analysis of extrapolating the Hall resistivity to the zero magnetic field to obtain the anomalous contribution becomes subtle, and so far only a few experiments and analyses have been performed in this regime. Clearly, futher experimental studies are required to clarify the scaling behaviors in this regime.

\section{Conclusions}\label{sec:conclusions}

In conclusions, we have developed a unified theory of anomalous Hall effect in ferromagnets, in terms of the fully quantum-mechanical transport theory for multi-band systems. It confirms that the anomalous Hall effect is determined by the intrinsic Berry-phase mechanism when (i) the Fermi level is located around an avoided-crossing of band dispersions in the momentum space, (ii) consequently the magnitude of $\sigma_{xy}$ is resonantly enhanced to the order of $e^2/(ha) \sim 10^3 \ \Omega^{-1}$ cm$^{-1}$, and (iii) the resistivity $\rho_{xx}$ is larger than $ (ha/e^2)(E_{\text{SO}}/E_F)\sim 1$-$10\ \mu\Omega\ {\rm cm}$ in 3$d$ transition-metals and $10$-$100\ \mu\Omega\ \text{cm}$ in 5$d$ and rare-earth compounds. In fact, the intrinsic contribution suffers from a partial cancellation due to the scattering events described as the vertex correction. Nevertheless, with these resonant conditions, it remains of the order of $e^2/h$ in two dimenions and $e^2/(ha)$ in three dimensions. Then, first-principles calculation can give a good estimate of $\sigma_{xy}$. By contrast, in the superclean systems with the lower resistivity, the skew scattering gives the leading contribution diverging in proportion to the quasiparticle lifetime. As the damping rate increases beyond the energy scale of spin-orbit interaction, the skew-scattering contribution gradually disappears and the anomalous Hall effect is dominated by the intirnsic contributions which are robust against the relaxation in the conducting regime. This extrinsic-intrinsic crossover needs to be verified by further careful experiments. For dirty systems with $\sigma_{xx}\lesssim e^2/ha$, another new scaling relation $\sigma_{xy}\propto\sigma_{xx}^{1.6}$ is obtained. Many ferromagnetic materials are located in this regime, and many experimental studies support this scaling relation. The present work resolves the long-standing puzzle and controversy on the mechanism of the anomalous Hall effect in a whole region of ferromagnetic metals at low temperatures, and reveals novel crossovers in quantum transport phenomena in multi-band systems.

\begin{acknowledgments}
The authors thank N.P. Ong, Y. Tokura, M. Kawasaki, Y. Onose, A. Asamitsu, S. Iguchi, A.H. MacDonald, J. Sinova, H. Fukuyama, and J. Inoue for useful dicussions and comments. S.O. was supported by Grant-in-Aids for Young Scientists (No. 19840053) from Japan Society for the Promotion of Science.
\end{acknowledgments}

\appendix

\section{Notations for detailed calculations}
\label{app:notation}

First, we introduce the following notations,
\begin{eqnarray}
  \hat{H}_0(\bm{p}) &=& \sum_{\mu=0,x,y,z}\hat{\sigma}^\mu H_0^\mu(\bm{p})
  \label{eq:H_0:4}\\
  \hat{G}^\alpha_b(\varepsilon,\bm{p}) &=& \sum_{\mu=0,x,y,z} \hat{\sigma}^\mu G_b^{\alpha\mu}(\varepsilon,\bm{p}),
  \label{eq:G:4}\\
  \hat{\Sigma}^{\alpha}_b(\varepsilon) &=& \sum_{\mu=0,x,y,z} \hat{\sigma}^\mu \Sigma^{\alpha\mu}_b(\varepsilon),
  \label{eq:Sigma:4}\\
  \hat{g}^{\alpha \mu}_b(\varepsilon)&=&\int\!\frac{d^2\bm{p}}{(2\pi\hbar)^2} G_b^{\alpha \mu}(\varepsilon,\bm{p}),
  \label{eq:g:4}
\end{eqnarray}
with $\alpha=R$, $A$, and $<$, and $b=0$ and ${E_y}$. Thoughout the Appendices, Greek symbols ($\mu$, $\nu$, $\rho$, $\cdots$) are used to label the components 0, $x$, $y$, and $z$, while italic ones ($i$, $j$, $\ell$, $\cdots$) are used only for $x$, $y$, and $z$. Contraction of these indices are implicitly assumed in all the expressions in the Appendices. 

\section{Explicit forms of $\hat{G}^R_0$ and $\hat{\Sigma}^R_0$}
\label{app:G^R_0}

Using the notations introduced in Appendix~\ref{app:notation}, it is ready to calculate the Green's functions and self-energies in the absence of the external fields from the coupled self-consistent equations~(\ref{eq:G^R,A:0}), (\ref{eq:Sigma^R,A:0}) and (\ref{eq:g^R,A:0}).
More explicitly, a straightforward calculation show
\begin{subequations}
\begin{eqnarray}
  \Sigma^{R0}_0(\varepsilon)&=&\frac{n_{\rm imp}u_{\text{imp}}\left(1-u_{\text{imp}}g^{R0}_0(\varepsilon)\right)}{(1-u_{\text{imp}}g^{R0}_0(\varepsilon))^2-u_{\text{imp}}^2g^{Rz}_0(\varepsilon)^2},
  \label{eq:Rashba:Sigma^R,A0_0}\\
  \Sigma^{Rz}_0(\varepsilon)&=&\frac{n_{\rm imp}u_{\text{imp}}^2g^{Rz}_0(\varepsilon)}{(1-u_{\text{imp}}g^{R0}_0(\varepsilon))^2-u_{\text{imp}}^2g^{Rz}_0(\varepsilon)^2},
  \label{eq:Rashba:Sigma^R,Az_0}\\
  \Sigma^{Rx}_0(\varepsilon)&=&\Sigma^{Ry}_0(\varepsilon)=0,
  \label{eq:Rashba:Sigma^R,Axy_0}\\
  g^{R0}_0(\varepsilon)&=&\frac{m}{4\pi\hbar^2}
  \sum_\sigma\log\frac{G^{R}_0(\varepsilon,\Lambda,\sigma)}{G^R_0(\varepsilon,0,\sigma)}
  -mv^2 \tilde{g}^{R}_0(\varepsilon),\ \ \ \ \ \ \ \ 
  \label{eq:Rashba:g^R,A0_0}\\
  g^{Rz}_0(\varepsilon)&=&(-\Delta_0+\Sigma^{Rz}_0(\varepsilon))\tilde{g}^{R}_0(\varepsilon)
  \label{eq:Rashba:g^R,Az_0}\\
  g^{Rx}_0(\varepsilon)&=&g^{Ry}_0(\varepsilon)=0
  \label{eq:Rashba:g^R,Axy_0}
\end{eqnarray}
\label{eq:Rashba:0}
\end{subequations}
with
\begin{widetext}
\begin{eqnarray}
  \tilde{g}^{R}_0(\varepsilon)&=&\frac{m}{4\pi\hbar^2 R^{R}(\varepsilon)}\left[
    \sum_\sigma\log\left(\varepsilon-p^2/2m+\mu-\Sigma^{R0}_0(\varepsilon)+mv^2+\sigma R^{R}(\varepsilon)\right)\right]_{p=0}^{p=\Lambda}
  \label{eq:Rashba:tg^R,A_0}\\
  G^{R}_0(\varepsilon,p,\pm)&=&\left(\varepsilon-p^2/2m+\mu-\Sigma^{R0}_0(\varepsilon)\mp\sqrt{v^2p^2+(-\Delta_0+\Sigma^{Rz}_0(\varepsilon))^2}\right)^{-1}
  \nonumber\\
  \label{eq:Rashba:G^R,A_0sigma}\\
  R^{R}(\varepsilon)&=&\sqrt{(mv^2)^2+2mv^2(\varepsilon+\mu-\Sigma^{R0}_0(\varepsilon))+(-\Delta_0+\Sigma^{Rz}_0(\varepsilon))^2}
  \label{eq:Rashba:F}
\end{eqnarray}
\end{widetext}
For later use, we also introduce $\tilde{G}^R_0$ through
\begin{eqnarray}
  G^{Ri}_0(\varepsilon,\bm{p})&=&\left(-v\epsilon_{ijz}p_j+\delta_{iz}(-M+\Sigma^{Rz}_0(\varepsilon))\right)\tilde{G}^R_0(\varepsilon,p),
  \nonumber\\
  \label{eq:tG^R_0}
\end{eqnarray}
with the fully anti-symmetric tensor $\epsilon_{ij\ell}$.

All the momentum integrations are performed analytically without any approximation as a function of $\varepsilon$, and then solve the self-consistent equations for each $\varepsilon$.

\section{Solving the self-consistent equations for $\hat{G}^<_{E_y,I}$ and $\hat{\Sigma}^<_{E_y,I}$ and calculation of $\sigma_{ij}^I$}
\label{app:sigma^I}

In the following, we show details in calculating the $E_y$-linear deviation of the self-energy due to the Fermi-surface contribution, $\hat{\Sigma}_{E_y,I}$, self-consistently with that of the Green's function, which are defined through Eqs.~(\ref{eq:Sigma^<:E}) and (\ref{eq:G^<:E}), respectively.
We exploit the notations defined by Eqs.~(\ref{eq:H_0:4}), (\ref{eq:G:4}), and (\ref{eq:Sigma:4}).
We start with Eq.~(\ref{eq:G^<:E,I}), whose component can be explicitly written by using $\Re G^{R\mu}=(G^{R\mu}+G^{A\mu})/2$ and $\Im G^{R\mu}=(G^{R\mu}-G^{A\mu})/2i$ as,
\begin{widetext}
\begin{subequations}
\begin{eqnarray}
  &&\Im\Sigma^{R\mu}_0(\varepsilon)G^{<\mu}_{E_y,I}(\varepsilon,\bm{p}) =
  \Im G^{R\mu}_0(\varepsilon,\bm{p})\Sigma^{<\mu}_{E_y,I}(\varepsilon)
  \nonumber\\
  &&
  +i\left((\Im\Sigma^{R\mu}_0(\varepsilon))(\bm{\nabla}_p \Re G^{R\mu}_0(\varepsilon,\bm{p}))-(\bm{\nabla}_p(H^\mu_0(\bm{p})+\Re\Sigma^{R\mu}_0(\varepsilon)))(\Im G^{R\mu}_0(\varepsilon,\bm{p}))\right),
  \nonumber\\
  \label{eq:G^<:E:I:QBE:0}\\
  &&\Im\vec{\Sigma}^R_0(\varepsilon)G^{<0}_{E_y,I}(\varepsilon,\bm{p})+\Im\Sigma^{R0}_0(\varepsilon)\vec{G}^<_{E_y,I}(\varepsilon,\bm{p})+(\vec{H}_0(\bm{p})+\Re\vec{\Sigma}^R_0(\bm{p}))\times\vec{G}^<_{E_y,I}(\varepsilon,\bm{p})
  \nonumber\\
  &=&(\Im \vec{G}^R_0(\varepsilon,\bm{p}))\Sigma^{<0}_{E_y,I}(\varepsilon)+(\Im G^{R0}_0(\varepsilon,\bm{p}))\vec{\Sigma}^<_{E_y,I}(\varepsilon)+\Re\vec{G}^R_0(\varepsilon,\bm{p})\times\vec{\Sigma}^<_{E_y,I}(\varepsilon)
  \nonumber\\
  &&{}+i\left((\Im\Sigma^{R0}_0(\varepsilon))(\bm{\nabla}_p \Re\vec{G}^R_0(\varepsilon,\bm{p}))+(\Im\vec{\Sigma}^R_0(\varepsilon))(\bm{\nabla}_p\Re G^{R0}_0(\varepsilon,\bm{p}))
  \right.\nonumber\\
  &&\left.\ \ \ \ {}-(\bm{\nabla}_p\vec{H}_0(\bm{p}))(\Im G^{R0}_0(\varepsilon,\bm{p}))-(\bm{\nabla}_pH^0_0(\bm{p}))(\Im\vec{G}^R_0(\varepsilon,\bm{p}))
  +(\Im\vec{\Sigma}^R_0(\varepsilon))\times(\bm{\nabla}_p\Im\vec{G}^R_0(\varepsilon,\bm{p}))\right).
  \label{eq:G^<:E:I:QBE:i}
\end{eqnarray}
\end{subequations}
These equations can be expressed in a matrix form
\begin{eqnarray}
  G^{<\mu}_{E_y,I}(\varepsilon,\bm{p})=\mathcal{L}^{<\mu\nu}_{0,I}(\varepsilon,\bm{p})\left(\mathcal{K}^{\nu\rho}_{0,I}(\varepsilon,\bm{p})\Sigma^{<\rho}_{E_y,I}(\varepsilon)+i\mathcal{K}^{\nu}_{E_y,I}(\varepsilon,\bm{p})\right)
  \label{eq:two-band:QBE:grad:E:<,I_mat}
\end{eqnarray}
with
\begin{eqnarray}
  {\mathcal{L}_{0,I}^{<-1}}^{\mu\nu}(\varepsilon,\bm{p})
  &=& \delta_{\mu\nu}\Im\Sigma^{R0  }_0(\varepsilon)
  +   \delta_{\mu0  }(1-\delta_{\nu0})\Im\Sigma^{R\nu}_0(\varepsilon)
  +   \delta_{\nu0  }(1-\delta_{\mu0})\Im\Sigma^{R\mu}_0(\varepsilon)
  -\epsilon_{\mu\nu\rho}(H^\rho_0(\bm{p})+\Re\Sigma^\rho_0(\varepsilon)),\ \ \ \ \ 
  \label{eq:L^<_0,E,I:inv}\\
  \mathcal{K}_{0,I}^{\mu\nu}(\varepsilon,\bm{p})&=&\delta_{\mu\nu}\Im G^{R0  }_0(\varepsilon,\bm{p})
  +   \delta_{\mu0  }(1-\delta_{\nu0})\Im G^{R\nu}_0(\varepsilon,\bm{p})
  +   \delta_{\nu0  }(1-\delta_{\mu0})\Im G^{R\mu}_0(\varepsilon,\bm{p})
  -\epsilon_{\mu\nu\rho}\Re G^{R\rho}_0(\varepsilon,\bm{p}),
  \label{eq:K^<_0,I}\\
  \mathcal{K}_{E_y,I}^0(\varepsilon,\bm{p})
  &=&-\Im\Sigma^{R0}_0(\varepsilon)(\bm{\nabla}_p\Re G^{R0}_0(\varepsilon,\bm{p}))+\Im\vec{\Sigma}^R_0(\varepsilon)(\bm{\nabla}_p\Re\vec{G}^R_0(\varepsilon,\bm{p}))
  \nonumber\\
  &&+(\bm{\nabla}_p H^0_0(\bm{p}))(\Im G^{R0}_0(\varepsilon,\bm{p}))
  -(\bm{\nabla}_p \vec{H}_0(\bm{p}))(\Im\vec{G}^R_0(\varepsilon,\bm{p}))
  \label{eq:K^<0_E,I}\\
  \mathcal{K}_{E_y,I}^i(\varepsilon,\bm{p})
  &=&\Im\Sigma^{R0}_0(\varepsilon)(\bm{\nabla}_p\Re G^{Ri}_0(\varepsilon,\bm{p}))
  -(\bm{\nabla}_p H^0_0(\bm{p}))(\Im G^{Ri}_0(\varepsilon,\bm{p}))
  \nonumber\\
  &&{}+ \Im\Sigma^{Ri}_0(\varepsilon)(\bm{\nabla}_p\Re G^{R0}_0(\varepsilon,\bm{p}))
  -(\bm{\nabla}_p H^i_0(\bm{p}))(\Im G^{R0}_0(\varepsilon,\bm{p}))
  \nonumber\\
  &&{}+\epsilon_{i\nu\rho}(\Im\Sigma^{R\nu}_0(\varepsilon))(\bm{\nabla}_p\Im G^{R\rho}_0(\varepsilon,\bm{p})),
  \label{eq:K^<i_E,I}
\end{eqnarray}
with the fully antisymmetric tensor $\epsilon_{\mu\nu\rho}$ that vanishes if any of $\mu$, $\nu$, and $\rho$ is 0.
Here, the equilibrium properties for the self-energy $\Sigma_0^{R\mu}$ with $\mu=0$, $x$, $y$, and $z$ have already been given in Eqs.~(\ref{eq:Rashba:Sigma^R,A0_0}), (\ref{eq:Rashba:Sigma^R,Az_0}), and (\ref{eq:Rashba:Sigma^R,Axy_0}), and hence those for the Green's function $G_0^{R\mu}$ can be obtained from the equilibrium Dyson equation. Then, Eqs.~(\ref{eq:L^<_0,E,I:inv}), (\ref{eq:K^<_0,I}), (\ref{eq:K^<0_E,I}), and (\ref{eq:K^<i_E,I}) are rewritten as
\begin{eqnarray}
  {\mathcal{L}_{0,I}^{<-1}}^{\mu\nu}(\varepsilon,\bm{p})
  &=&\delta_{\mu\nu}\Im\Sigma^{R0}_0(\varepsilon)+\left(\delta_{\mu0}\delta_{\nu z}+\delta_{\mu z}\delta_{\nu0}\right)\Im\Sigma^{Rz}_0(\varepsilon)-\epsilon_{\mu\nu z}(-M+\Re\Sigma^{Rz}_0(\varepsilon))-v(\delta_{\mu0}p_\nu-\delta_{\mu z}p_\nu),
  \label{eq:Rashba:L^<_0,E,I:inv}\\
  \mathcal{K}_{0,I}^{\mu\nu}(\varepsilon,\bm{p})
  &=&\delta_{\mu\nu}\Im G^{R0}_0(\varepsilon,\bm{p})+(\delta_{\mu0}\delta_{\nu z}+\delta_{\mu z}\delta_{\nu0})\Im G^{Rz}_0(\varepsilon,\bm{p})-\epsilon_{\mu\nu i}G^{Ri}_0(\varepsilon,\bm{p}),
  \label{eq:Rashba:K^<_0,I}\\
  \mathcal{K}_{E_j,I}^0(\varepsilon,\bm{p})
  &=&-\Im\Sigma^{R0}_0(\varepsilon)(\partial_{p_j}\Re G^{R0}_0(\varepsilon,\bm{p}))+\Im\Sigma^{Rz}_0(\varepsilon)(\partial_{p_j}\Re G^{Rz}_0(\varepsilon,\bm{p}))
  \nonumber\\
  &&+(p_j/m)(\Im G^{R0}_0(\varepsilon,p))+v\epsilon_{ijz}(\Im G^{Ri}_0(\varepsilon,\bm{p})),
  \label{eq:Rashba:K^<0_E,I}\\
  \mathcal{K}_{E_j,I}^i(\varepsilon,\bm{p})
  &=&\Im\Sigma^{R0}_0(\varepsilon)(\partial_{p_j}\Re G^{Ri}_0(\varepsilon,\bm{p}))
  -(p_j/m)(\Im G^{Ri}_0(\varepsilon,\bm{p}))
  \nonumber\\
  &&{}+ \Im\Sigma^{Ri}_0(\varepsilon)(\partial_{p_j}\Re G^{R0}_0(\varepsilon,\bm{p}))
  +v\epsilon_{ijz}(\Im G^{R0}_0(\varepsilon,\bm{p}))
  -\epsilon_{i\ell z}(\Im\Sigma^{Rz}_0(\varepsilon))(\partial_{p_j}\Im G^{R\ell}_0(\varepsilon,\bm{p})),
  \label{eq:Rashba:K^<i_E,I}
\end{eqnarray}

On the other hand, from Eqs.~(\ref{eq:g^R,A:0}), (\ref{eq:G:4}), (\ref{eq:g:4}), and (\ref{eq:Rashba:g^R,Axy_0}), the $T$-matrix approximation to $\hat{\Sigma}^<_{E_y,I}$, which has been formally described in Eq.~(\ref{eq:Sigma^<:E,I}), is reduced to the following expression in the matrix-form representation,
\begin{eqnarray}
  \Sigma^{<\mu}_{E_y,I}(\varepsilon) &=& \mathcal{B}^{<\mu\nu}_{0,I}(\varepsilon)g^{<\nu}_{E_y,I}(\varepsilon),
  \label{eq:Sigma^<:E,I-B}\\
  \mathcal{B}^{<\mu\nu}_{0,I}(\varepsilon) 
  &=& n_{\text{imp}}u_{\text{imp}}\Biggl[
    \delta_{\mu\nu}\frac{\left|1-u_{\text{imp}}g^{R0}_0(\varepsilon)\right|^2+\left|u_{\text{imp}}g^{Rz}_0(\varepsilon)\right|^2}{\left|(1-u_{\text{imp}}g^{R0}_0(\varepsilon))^2-u_{\text{imp}}^2g^{Rz}_0(\varepsilon)^2\right|^2}
  \nonumber\\
  &&+(\delta_{\mu0}\delta_{\nu z}+\delta_{\mu z}\delta_{\nu 0})
  \frac{2\Re\left((1-u_{\text{imp}}g^{R0}_0(\varepsilon))u_{\text{imp}}g^{Az}_0(\varepsilon)\right)}{\left|(1-u_{\text{imp}}g^{R0}_0(\varepsilon))^2-u_{\text{imp}}^2g^{Rz}_0(\varepsilon)^2\right|^2}
  -\epsilon_{\mu\nu z}\frac{2\Im\left((1-u_{\text{imp}}g^{R0}_0(\varepsilon))u_{\text{imp}}g^{Az}_0(\varepsilon)\right)}{\left|(1-u_{\text{imp}}g^{R0}_0(\varepsilon))^2-u_{\text{imp}}^2g^{Rz}_0(\varepsilon)^2\right|^2}\Biggr].
  \ \ \ \ \ \ \ 
  \label{eq:B^<:E,I}
\end{eqnarray}
\end{widetext}
Substituting Eq.~(\ref{eq:Sigma^<:E,I-B}) into Eq.~(\ref{eq:two-band:QBE:grad:E:<,I_mat}) and integrating over $\bm{p}$, we obtain
\begin{eqnarray}
  g^{<\mu}_{E_y,I}(\varepsilon)&=&i\mathcal{C}^{<\mu\nu}_{E_y,I}(\varepsilon)\int\!\!\!\frac{d^2\bm{p}}{(2\pi\hbar)^2}\mathcal{L}^{<\nu\rho}_{0,I}(\varepsilon,\bm{p})\mathcal{K}^{\rho}_{0,I}(\varepsilon,\bm{p}),
  \label{eq:g^<}\\
  \mathcal{C}^{<-1}_{E_y,I}{}^{\mu\nu}(\varepsilon)&=&\delta_{\mu\nu}-\int\!\!\!\frac{d^2\bm{p}}{(2\pi\hbar)^2}\mathcal{L}^{<\mu\rho}_{0,I}(\varepsilon,\bm{p})\mathcal{K}^{\rho\sigma}_{0,E_y}(\varepsilon,\bm{p})\mathcal{B}^{<\sigma\nu}_{0,I}(\varepsilon).
  \nonumber\\
  \label{eq:C}
\end{eqnarray}
All the momentum integrations are performed analytically without any approximation to avoid an unnecessary instability which otherwize occurs in the numerical momentum integration, and then the $\varepsilon$ dependences are self-consistently calculated by numerical iteractions.

Finally, $\sigma_{ij}^I$ is calculated through Eq.~(\ref{eq:sigma^I}), which is rewritten as
\begin{eqnarray}
  \sigma_{ij}^I &=& -\frac{e^2\hbar}{\pi i}\left[\int\!\!\frac{d^2\bm{p}}{(2\pi\hbar)^2}
  \frac{p_i}{m}G^{<0}_{E_j,I}(\mu,\bm{p})+v\epsilon_{i\ell z}g^{<\ell}_{E_j,I}(\mu)\right],
  \nonumber\\
  \label{eq:Rashba:sigma^I}
\end{eqnarray}

\section{Solving the self-consistent equations for $\hat{G}^R_{E_y}$ and $\hat{\Sigma}^R_{E_y}$ and calculation of $\sigma_{ij}^{II}$}
\label{app:sigma^II}

The quantum contribution to the linear response $\hat{G}^<_{E_y,II}$ can be calculated from the retarded (advanced) Green's functions $\hat{G}^{R}_{E_y}(\varepsilon)$ given by Eq.~(\ref{eq:G^R,A:E}). In the following, the integral equations for them are explicitly derived. We also employ the notations defined by Eqs.~(\ref{eq:H_0:4}), (\ref{eq:G:4}), and (\ref{eq:Sigma:4}).

First, each component of Eq.~(\ref{eq:G^R,A:E}) is expressed as
\begin{widetext}
\begin{subequations}
  \begin{eqnarray}
    G^{R0}_{E_y}(\varepsilon,\bm{p})
    &=&(G^{R0}_0(\varepsilon,\bm{p})^2+\vec{G}^{R}_0(\varepsilon,\bm{p})^2)\Sigma^{R0}_{E_y}(\varepsilon)+2G^{R0}_0(\varepsilon) \vec{G}^{R}_0(\varepsilon,\bm{p})\cdot\vec{\Sigma}^{R}_{E_y}(\varepsilon)
    \nonumber\\
    &&{}-(G^{R0}_0(\varepsilon,\bm{p})^2-\vec{G}^{R}_0(\varepsilon,\bm{p})^2)(\partial_{\varepsilon}\vec{\Sigma}^{R}_0(\varepsilon))\cdot \vec{G}^{R}_0(\varepsilon,\bm{p})\times(\bm{\nabla}_p\vec{H}_0(\bm{p}))
  \label{eq:two-band:G^<R,A0:E}\\
  \vec{G}^{R}_{E_y}(\varepsilon,\bm{p})
  &=&(G^{R0}_0(\varepsilon,\bm{p})^2-\vec{G}^{R}_0(\varepsilon,\bm{p})^2)
  \left(\vec{\Sigma}^{R}_{E_y}(\varepsilon)-(\partial_{\varepsilon}\vec{\Sigma}^{R}_0(\varepsilon))\times\vec{G}^{R}_0(\varepsilon,\bm{p})(\bm{\nabla}_pH^0_0(\bm{p}))\right.
  \nonumber\\
  &&\left.\ \ \ {}+(1-\partial_{\varepsilon}\Sigma^{R0}_0(\varepsilon))\vec{G}^{R}_0(\varepsilon,\bm{p})\times(\bm{\nabla}_p\vec{H}_0(\bm{p}))
  -G^{R0}_0(\varepsilon,\bm{p})(\partial_{\varepsilon}\vec{\Sigma}^{R}_0(\varepsilon))\times(\bm{\nabla}_p\vec{H}_0(\bm{p})\right)
  \nonumber\\
  &&{}+2\vec{G}^{R}_0(\varepsilon,\bm{p})\left(G^{R0}_0(\varepsilon,\bm{p})\Sigma^{R0}_{E_y}(\varepsilon)+\vec{G}^{R}_0(\varepsilon,\bm{p})\vec{\Sigma}^{R}_{E_y}(\varepsilon)\right)
  \label{eq:two-band:G^R,Ai:E}
  \end{eqnarray}
  \label{eq:two-band:G^R,A:E}
\end{subequations}
Applying the above formula to the present model, we obtain
\begin{subequations}
\begin{eqnarray}
  G^{R0}_{E_y}(\varepsilon,\bm{p})
  &=&(G^{R0}_0(\varepsilon,\bm{p})^2+\vec{G}^{R}_0(\varepsilon,\bm{p})^2)\Sigma^{R0}_{E_y}(\varepsilon)
  +2G^{R0}_0(\varepsilon,\bm{p})\vec{G}^{R}_0(\varepsilon,\bm{p})\vec{\Sigma}^{R}_{E_y}(\varepsilon)
  -v^2\bm{z}\times\bm{p}(\partial_\varepsilon\Sigma^{Rz}_0(\varepsilon))\tilde{G}^{R}_0(\varepsilon,\bm{p})^2
  \label{eq:G^R,A0:E}\\
  G^{Rz}_{E_y}(\varepsilon,\bm{p})
  &=&\tilde{G}^{R}_0(\varepsilon,\bm{p})\Sigma^{Rz}_{E_y}(\varepsilon)
  +2G^{Rz}_0(\varepsilon,\bm{p})(G^{R0}_0(\varepsilon,\bm{p})\Sigma^{R0}_{E_y}(\varepsilon)+\vec{G}^{R}_0(\varepsilon,\bm{p})\vec{\Sigma}^{R}_{E_y}(\varepsilon))
  +v^2\bm{z}\times\bm{p}(1-\partial_\varepsilon\Sigma^{R0}_0(\varepsilon))\tilde{G}^{R}_0(\varepsilon,\bm{p})^2
  \nonumber\\
  \label{eq:Rashba:G^R,Az:E}\\
  G^{Ri}_{E_j}(\varepsilon,\bm{p})
  &=&\tilde{G}^{R}_0(\varepsilon,\bm{p})\Sigma^{Ri}_{E_j}(\varepsilon)
  +2G^{Ri}_0(\varepsilon,\bm{p})(G^{R0}_0(\varepsilon,\bm{p})\Sigma^{R0}_{E_j}(\varepsilon)+\vec{G}^{R}_0(\varepsilon,\bm{p})\vec{\Sigma}^{R}_{E_j}(\varepsilon))
  \nonumber\\
  &&{}+v\left(\delta_{ij}((1-\partial_\varepsilon\Sigma^{R0}_0(\varepsilon))G^{Rz}_0(\varepsilon,\bm{p})-(\partial_\varepsilon\Sigma^{Rz}_0(\varepsilon))G^{R0}_0(\varepsilon,\bm{p}))
  -\frac{p_ip_j}{m}(\partial_\varepsilon\Sigma^{Rz}_0(\varepsilon))\tilde{G}^{R}_0(\varepsilon,\bm{p})\right)\tilde{G}^{R}_0(\varepsilon,\bm{p})
  \label{eq:Rashba:G^R,Ai:E}
\end{eqnarray}
\label{eq:Rashba:G^R,A:E}
\end{subequations}
Using Eqs.~(\ref{eq:g^R,A:0}) and (\ref{eq:Rashba:g^R,Axy_0}), components of the self-energy $\hat{\Sigma}^{R}_{E_y}(\varepsilon)$ given by Eq.~(\ref{eq:Sigma^R,A:E}) can be explicitly written as
\begin{subequations}
\begin{eqnarray}
  \Sigma^{R0}_{E_y}(\varepsilon) &=& n_{\text{imp}}u_{\text{imp}}^2\left((1-u_{\text{imp}}g^{R0}_0(\varepsilon))^2-u_{\text{imp}}^2g^{Rz}_0(\varepsilon)^2\right)^{-2}
  \nonumber\\
  &&\times\left[\left((1-u_{\text{imp}}g^{R0}_0(\varepsilon))^2+u_{\text{imp}}^2g^{Rz}_0(\varepsilon)^2\right)g^{R0}_{E_y}(\varepsilon)
    +2(1-u_{\text{imp}}g^{R0}_0(\varepsilon))g^{Rz}_0(\varepsilon)g^{Rz}_{E_y}(\varepsilon)\right],
  \label{eq:Rashba:Born_all:Sigma^R,A:E:0}\\
  \Sigma^{Rz}_{E_y}(\varepsilon) &=& n_{\text{imp}}u_{\text{imp}}^2\left((1-u_{\text{imp}}g^{R0}_0(\varepsilon))^2-u_{\text{imp}}^2g^{Rz}_0(\varepsilon)^2\right)^{-2}
  \nonumber\\
  &&\times\left[\left((1-u_{\text{imp}}g^{R0}_0(\varepsilon))^2+u_{\text{imp}}^2g^{Rz}_0(\varepsilon)^2\right)g^{Rz}_{E_y}(\varepsilon)
     +2(1-u_{\text{imp}}g^{R0}_0(\varepsilon))g^{Rz}_0(\varepsilon)g^{R0}_{E_y}(\varepsilon)\right],
  \label{eq:Rashba:Born_all:Sigma^R,A:E:z}\\
  \Sigma^{Ri}_{E_y}(\varepsilon) &=& n_{\text{imp}}u_{\text{imp}}^2\left((1-u_{\text{imp}}g^{R0}_0(\varepsilon))^2-u_{\text{imp}}^2g^{Rz}_0(\varepsilon)^2\right)^{-1}g^{Ri}_{E_y}(\varepsilon).
  \label{eq:Rashba:Born_all:Sigma^R,A:E:i}
\end{eqnarray}
\label{eq:Rashba:Born_all:Sigma^R,A:E}
\end{subequations}
Integrating Eqs.~(\ref{eq:Rashba:G^R,A:E}) over $\bm{p}$ and taking into account Eqs.~(\ref{eq:Rashba:0}) and (\ref{eq:Rashba:Born_all:Sigma^R,A:E}), it is found that 
\begin{subequations}
\begin{eqnarray}
  g^{R0}_{E_y}(\varepsilon)&=& g^{Rz}_{E_y}(\varepsilon)=0,
  \label{eq:Rashba:g^R,A0,z:E}\\
  g^{Ri}_{E_j}(\varepsilon)
  &=&\Sigma^{Ri}_{E_j}(\varepsilon)\int\!\frac{d^2\bm{p}}{(2\pi\hbar)^2}\,(1+v^2p^2\tilde{G}^{R}_0(\varepsilon,p))\tilde{G}^{R}_0(\varepsilon,p)
  \nonumber\\
  &&{}+v\delta_{ij}\int\!\frac{d^2\bm{p}}{(2\pi\hbar)^2}\,\tilde{G}^{R}_0(\varepsilon,p)^2\left[(1-\partial_\varepsilon\Sigma^{R0}_0(\varepsilon))(-\Delta_0+\Sigma^{Rz}_0(\varepsilon))
    -(\varepsilon+\mu-\Sigma^{R0}_0(\varepsilon))(\partial_\varepsilon\Sigma^{Rz}_0(\varepsilon))\right]
  \nonumber\\
  &=&v\delta_{ij}\int\!\frac{d^2\bm{p}}{(2\pi\hbar)^2}\,\tilde{G}^{R}_0(\varepsilon,p)^2\left[(1-\partial_\varepsilon\Sigma^{R0}_0(\varepsilon))(-\Delta_0+\Sigma^{Rz}_0(\varepsilon))
    -(\varepsilon+\mu-\Sigma^{R0}_0(\varepsilon))(\partial_\varepsilon\Sigma^{Rz}_0(\varepsilon))\right]
  \nonumber\\
  &&{}\times\left[1-n_{\rm imp}u_{\text{imp}}^2\left((1-u_{\text{imp}}g^{R0}_0(\varepsilon))^2-u_{\text{imp}}^2g^{Rz}_0(\varepsilon)^2\right)^{-1}
    \int\!\frac{d^2\bm{p}}{(2\pi\hbar)^2}\,(1+v^2p^2\tilde{G}^{R}_0(\varepsilon,p))\tilde{G}^{R}_0(\varepsilon,p)\right]^{-1},
  \label{eq:Rashba:g^R,Ai:E}\\
  \Sigma^{R0}_{E_y}(\varepsilon)&=&\Sigma^{Rz}_{E_y}(\varepsilon)=0,
  \label{eq:Rashba:Sigma^R,A0,z:E}
\end{eqnarray}
\end{subequations}
with $i$ and $j$ being $x$ or $y$. Then, we obtain
\begin{eqnarray}
  G^{R0}_{E_j}(\varepsilon,\bm{p})&=&-v\epsilon_{ijz}p_i\left(2G^{R0}_0(\varepsilon,\bm{p})\Sigma^{Rj}_{E_j}(\varepsilon)+v(\partial_\varepsilon\Sigma^{Rz}_0(\varepsilon))\tilde{G}^{R}_0(\varepsilon,p)\right)\tilde{G}^{R}_0(\varepsilon,p).
  \label{eq:Rashba:G^R,A0:E}
\end{eqnarray}

Finally, we obtain
\begin{eqnarray}
  \sigma_{xy}^{II}&=&-2e^2\hbar\int\!\frac{d\varepsilon}{\pi}f(\varepsilon)\int\!\frac{d^2\bm{p}}{(2\pi \hbar)^2}\Im\left(\frac{p_x}{m} G^{R0}_{E_y}(\varepsilon,\bm{p})+v G^{Ry}_{E_y}(\varepsilon,\bm{p})\right)
  \nonumber\\
  &=& 2e^2\hbar v^2\Im\int\!\frac{d\varepsilon}{\pi}f(\varepsilon)\left[(1-\partial_\varepsilon\Sigma^{R0}_0(\varepsilon))(-\Delta_0+\Sigma^{Rz}_0(\varepsilon))
    -(\varepsilon+\mu-\Sigma^{R0}_0(\varepsilon))(\partial_\varepsilon\Sigma^{Rz}_0(\varepsilon))\right]
  \nonumber\\
  &&\times\int\!\frac{d^2\bm{p}}{(2\pi\hbar)^2}\tilde{G}^R_0(\varepsilon,p)^2\left[1+n_{\text{imp}}u_{\text{imp}}^2\left((1-u_{\text{imp}}g^{R0}_0(\varepsilon))^2-u_{\text{imp}}^2g^{Rz}_0(\varepsilon)^2\right)^{-1}2\int\!\frac{d^2\bm{p}}{(2\pi\hbar)^2}\frac{p^2}{2m}G^{R0}_0(\varepsilon,p)\tilde{G}^R_0(\varepsilon,p)\right]
  \nonumber\\
  &&\times\left[1-n_{\rm imp}u_{\text{imp}}^2\left((1-u_{\text{imp}}g^{R0}_0(\varepsilon))^2-u_{\text{imp}}^2g^{Rz}_0(\varepsilon)^2\right)^{-1}
    \int\!\frac{d^2\bm{p}}{(2\pi\hbar)^2}\,(1+v^2p^2\tilde{G}^{R}_0(\varepsilon,p))\tilde{G}^{R}_0(\varepsilon,p)\right]^{-1}
  \nonumber\\
  &&+2e^2\hbar v^2\Im\int\!\frac{d\varepsilon}{\pi}f(\varepsilon)(\partial_\varepsilon\Sigma^{Rz}_0(\varepsilon))\int\!\frac{d^2\bm{p}}{(2\pi\hbar)^2}\frac{p^2}{2m}\tilde{G}^R_0(\varepsilon,p)^2,
  \label{eq:Rashba:sigma^xy_II}
\end{eqnarray}
\end{widetext}
and a trivial result of $\sigma_{xx}^{II}=0$. Note that in the limit of a large momentum cutoff, the square brackets appearing in the numerator and the denominator of the first term of Eq.~(\ref{eq:Rashba:sigma^xy_II}) coincide, and hence the vertex correction or equivalently the effect of $\hat{\Sigma}^{R,A}_{E_y}$ is canceled out for $\sigma_{xy}^{II}$.
Again, all the momentum integrations are performed analytically and a special care should be taken in confirming a convergence of the numerical integration over $\epsilon$ for $\sigma_{xy}^{II}$.

\end{document}